\documentclass[letterpaper,12pt,leqno]{article}
\usepackage{paper}
\hypersetup{pdftitle={crypto_emp_ap}}
\newcommand{\bib}{bibliography.bib}

\newcommand{\mycomment}[1]{}

\begin{document}

\title{Empirical Crypto Asset Pricing}
\author{Adam Baybutt
    \thanks{\textit{Contact}: adam baybutt at protonmail dot com. \textit{Acknowledgments}: I thank the following people for valuable comments and suggestions: Jesper B\"ojeryd, Mikhail Chernov, Denis Chetverikov, Valentin Haddad, Basil Halperin, Manu Navjeevan, Andres Santos, and Pierre-Olivier Weill. I thank the following individuals for their research assistance: Jacob Brophy, Chuhan Guo, and Tiffany Zho. Thanks to Pascal Michaillat for formatting. Thank you to Coin Metrics, CoinMarketcap, and Glassnode for providing academic research discounts on data purchases. Replication code available at \url{https://github.com/adambaybutt/crypto_asset_pricing}.
    }}
\date{Last updated: \today}
\begin{titlepage}\maketitle

We motivate the study of the crypto asset class with eleven empirical facts, and study the drivers of crypto asset returns through the lens of univariate factors. We argue crypto assets are a new, attractive, and independent asset class. In a novel and rigorously built panel of crypto assets, we examine pricing ability of sixty three asset characteristics to find rich signal content across the characteristics and at several future horizons. Only univariate financial factors (i.e., functions of previous returns) were associated with statistically significant long-short strategies, suggestive of speculatively driven returns as opposed to more fundamental pricing factors.

\end{titlepage}

\newpage
\section{Introduction}\label{s:introduction}

We motivate the study of the crypto asset class with eleven empirical facts and investigate the drivers of crypto asset returns through the lens of univariate factors.

\paragraph{Why Crypto} \cite{nakamoto2008bitcoin} gifted a novel mechanism design known as Proof-of-Work, enabling a set of adversarial entities to reach consensus on the current state of an open database using cryptography, often framed as a solution to the Byzantine Generals' Problem \citep{lamport1982byzantine}. The Bitcoin blockchain launched in early 2009, employing Proof-of-Work to pioneer a censorship-resistant digital transaction ledger. This innovation introduced a permissionless payment network for transferring bitcoin, its native digital asset. The emergence of Nakamoto Consensus, along with other blockchain-based consensus mechanisms that followed, enabled the scarcity of digital information, particularly in the form of digital or crypto assets, and thus introduced a new area of economic research.

\paragraph{A New, Attractive, and Independent Asset Class} We motivate research into the return dynamics of these crypto assets by establishing the following empirical facts in Section \ref{s:empfacts}. The advent of Bitcoin sparked a Cambrian explosion of other crypto assets, evolving from initial valuations as collectibles into a trillion dollar asset class. Bitcoin has matured into a substantial payment network, settling hundreds of billions of dollars annually, with the large majority of transactions settling for a cost of less than one dollar, thereby offering monetary functions with distributed consensus. 

Bitcoin exhibited superior risk-adjusted returns when compared to traditional asset classes over our study period of 2018-2022, inclusive. With regard to independence, bitcoin has lower correlations with the Nasdaq and the S\&P500, at 0.23 and 0.21 respectively, as compared to gold's correlations with these indices at 0.26 and 0.28. Moreover, bitcoin's correlation with other assets exhibits significant temporal variance, including several quarters of zero or negative correlation with the Nasdaq; their high correlation (> 0.3) is only a recent phenomenon seen in 2022. While these measures are suggestive of an independent asset class, a possibly sufficient statistic is whether there are risk-adjusted return gains from including crypto assets in one's portfolio. From diversifying a risk portfolio of holding 100\% Nasdaq to instead holding 60\% Nasdaq and 40\% the crypto market, one would obtain a Sharpe Ratio gain of 0.53 (from 0.43 to 0.96).

\paragraph{Crypto Signals} The emergence of hundreds of crypto assets expands to a new asset class the central focus of empirical asset pricing: the search for explanations of why different assets earn different average returns. A fundamentally unique aspect of the crypto asset class is open state: the state of the digital ledger is readable. This is termed onchain data where one has access to the (onchain) economy's full history of transactions. For instance, we directly observe the hodling time of all Bitcoin wallets to discover a majority of wallets utilize bitcoin as a store of value rather than for speculatively trading.

In this manuscript, we formulate several novel crypto asset characteristics in addition to investigating the signal content of characteristics previously studied in the literature. An additional distinction of this study is to build a panel of tradable crypto asset excess return data with more realistic inclusion criteria than previously studied in the literature. In examining the signal content across this rich set of asset characteristics, although there are some redundant characteristics and signal decays over the study years, we observe numerous sources of signal for the cross-section of one-week-ahead expected returns. 

\paragraph{Empirical Setting} To assemble a weekly panel of tradable crypto assets, we prospectively identify, at the start of each month from 2018 to 2022, inclusive, tradable crypto assets on US centralized crypto exchanges with sufficient trading volume and market capitalization, which results in the number of assets growing from 10 on January 7, 2018 to 204 on December 1, 2022. There are 210 unique assets in the panel.

Motivated by a one percent threshold on an order book's volume, the most restrictive inclusion criteria applied each month, among several other criteria, is for each asset to have a median weekly volume across US exchanges of \$500k over the trailing three months. Using this strict set of inclusion criteria to study tradable assets without bias, our panel has a challenge wherein assets repeatedly enter and leave the panel over time as they rise above and fall below the inclusion criteria. We thus have to reform the panel monthly when fitting models.

This panel of weekly crypto asset excess returns is not only novel to the literature based on the inclusion criteria but also given it contains several novel asset characteristics across the sixty three characteristics studied within the following categories: onchain, social, financial, momentum, exchange, and microstructure. We do note that five years of data is limited, but this is the current state for empirical crypto asset pricing.

\paragraph{Univariate Factor Models} In studying sixty three univariate factors formed as the long-short quintile portfolios sorted on each asset characteristic, we find six financial factors (e.g., three momentums, beta, idiosyncratic skewness, and 5\% shortfall) are the only ones associated with significant differences in average one-week-ahead excess returns. Across the thirty quintile portfolios, twenty four followed a monotonic pattern within characteristic, with only the two week momentum having monotonic quintile portfolios. The time-series average weekly excess return spread (and annualized Sharpe ratios) for the zero-investment long-short strategies are 1.5\% (0.78) for two week momentum, 1.2\% (0.75) for one month industry momentum, 1.4\% (0.87) for two month industry momentum, 1.6\% (0.76) for one week beta, 1.2\% (0.79) for one month idiosyncratic skewness, and 1.4\% (0.76) for one week 5\% expected shortfall.

It is of note that all of these statistically significant strategies were formed on functions of previous returns, that is, momentum and financial characteristics. This is suggestive that crypto asset returns are not driven by fundamental factors. Although capturing differences in industries lead to significant long momentum strategies, more fundamental pricing characteristics were not associated with significant spreads between top and bottom quintile portfolios. With this more rigorous panel, many of the univariate factor results established in the existing literature fail to replicate. Moreover, as discussed in Section \ref{s:uni_factor}, many characteristics have economically significant return spreads and Sharpe ratios, although the short panel of five years and the volatility of returns in this asset class suppress the resulting \textit{t}-statistics.

\paragraph{Relevant Literature} This paper builds on the nascent empirical crypto asset pricing literature. Although a recent literature, there are dozens papers, of which we will discuss a small selection, studying the asset class and the performance of univariate factor sorts, which we replicate with a more realistic panel of crypto assets. \cite{shams2020structure} studies univariate factor performance in a panel of crypto assets, including novel social media and investor base measures. \cite{liu2022common} studies the cross sectional pricing ability of univariate factor portfolio sorts and multivariate observable (i.e., market, size, and momentum) factor model. Several univariate factors have statistically significant results, including characteristics formed using market capitalization, previous prices, momentums, and volumes. The three-factor model accounts for the performance of these univariate strategies. 

Next this paper also contributes novel empirical facts, including motivating crypto as a new asset class, presenting bitcoin's use as a store of value and payment network, and discussing the pricing ability of several new asset characteristics. \cite{makarov2020trading} provide empirical evidence of arbitrage opportunities between centralized exchanges, in particular across countries as bitcoin often trades at a premium outside the United States. \cite{hu2019cryptocurrencies} present a set of stylized facts on crypto as a new investable instrument, including crypto's low correlation with gold and equities but significant covariance with bitcoin, which is perhaps driven by the need to purchase bitcoin to access altcoins. \cite{borri2019conditional} estimate conditional value at risk in a small number of crypto assets to show these crypto assets are highly exposed to tail-risk within the crypto asset class, but not exposed to tail-risk with respect to traditional asset classes. Additionally, the authors present mean variance gains from a small crypto allocation to a traditional portfolio. \cite{bianchi2020cryptocurrencies} corroborates these findings showing crypto exhibits a low correlation to traditional asset classes as well as presenting results on previous volumes predicting crypto asset returns. \cite{liu2021risks} discover momentum and proxies for investor attention explain variation in crypto asset returns, yet major crypto assets do not comove with traditional asset prices nor macroeconomic factors. \cite{zhang2020idiosyncratic}, \cite{zhang2021downside}, and \cite{zhang2023liquidity} present significant risk premia results in a panel of crypto assets for idiosyncratic volatility, downside risk, and liquidity. \cite{bianchi2022dynamics} finds investor attention significantly predicts crypto asset returns. \cite{cheah2022predictability} finds several asset characteristics and macroeconomic factors with significant predictability for bitcoin returns, while stock and bond market common factors have no correlation with bitcoin returns.

As a final contribution to the empirical crypto asset pricing literature, this paper identifies several improvements for building a realistic panel of tradable crypto assets. For context, there are low barriers to entry in launching a new crypto asset, resulting in tens of thousands of crypto assets with a market as of 2023. However, the vast majority have insufficient liquidity for empirical study given trivial volume would have significant price impact. Moreover, many are not available on a US exchanges. Thus, as we will argue in Section \ref{s:descdata} that the literature is currently studying asset returns which were not tradable. Our criteria suggests a tradable panel starting with ten crypto assets in 2018 expanding to about two hundred assets at the end of 2022. However, by way of example, in similar time periods, \cite{liebi2022there} studies 652 cryptoassets; \cite{borri2022crypto} measures risk premia of observable factors using the method of \cite{giglio2021asset} in a panel of about seven hundred assets; and, \cite{cong2022value} studies risk premiums and factor pricing ability in a panel of four thousand crypto assets.

\section{Description of Data}\label{s:descdata}

\paragraph{Panel Overview} We obtain hourly crypto asset prices, trading volumes, and market capitalization from Coin Metrics and CoinAPI for 2018-2022, inclusive.\footnote{It should be noted that market capitalization in crypto is difficult to measure given the various measures of token supply. We use Coin Metrics' measure which aims to estimate the free float supply; see \url{https://coinmetrics.io/free-float-supply-a-better-measure-of-market-capitalization/} for further discussion.} Specifically, we collect these data for all assets with at least four months of trading history and a spot market to USD, USDC, or USDT on one of the following United States exchanges.\footnote{Specifically, : Binance US, Bitstamp, Coinbase, Crypto.com, FTX US, Gemini, Kraken, and Kucoin.} We exclude years before 2018 given a very small number of relevant assets, and we save 2023 for future out of sample experiments. Price is the volume-weighted average hourly candle mid-price across these exchanges where we average all variables first across the two data providers. \footnote{To our knowledge, this is a uncommon approach in the literature to use the actually tradable price as opposed to a global price across all exchanges. The mean absolute error between the global price uses the volume-weighted average hourly price and the actual tradable price from the US exchanges is \$1.71 while the mean absolute error in weekly return is 95 bps.} Excess returns are formed from this tradable price measure after removing the one month Treasury-bill rate to proxy for the risk-free rate. Our weekly panel is aggregated from this hourly panel as we will discuss. 

\paragraph{Rolling Inclusion Criteria} Crypto assets have a challenging empirical asset pricing problem of how to define a relevant temporarly-changing asset universe. To illustrate, note CoinMarketCap listed the market price of 1381 crypto assets on January 2, 2018. \footnote{\url{https://web.archive.org/web/20180102053542/https://coinmarketcap.com/all/views/all/}.} However, the vast majority of these assets are not tradable from the view of empirically studying their returns. That is, as we will discuss in the next paragraph, the tradable volume on US exchanges is only sufficiently high for a small number of assets, such that, historical returns will not be distorted by price impact. Moreover, there is volatility in the trading volume so one must sequentially build an asset universe.

On the first day of each study month, we define an asset as tradable in our study if it has twelve weeks of trailing data to form relevant characteristics; it has a spot market on a US exchange to USD, USDC, or USDT; it is not a stablecoin or synthetic asset (e.g., wrapped BTC, PAXG, etc.); its average market capitalization over the trailing three months was above one basis point of the total crypto market capitalization\footnote{Many other papers define a static market capitalization threshold of \$1MM, which will not be responsive to the significant expansion and contraction of market capitalizations in the asset class, e.g., \cite{liu2022common}, \cite{liu2021risks}, \cite{shams2020structure}, \cite{cong2022value}, etc.}; it has nonzero trading volume on all trailing twelve weeks; and, finally, its median weekly trading volume is above \$500,000 in the trailing twelve weeks.\footnote{To our knowledge, this is a novel approach in the empirical crypto asset pricing literature where we use a heuristic of staying below 1\% of the volume to assume no price impact and a weekly trade size for a given asset of \$5,000. See transaction cost discussion in Section \ref{s:uni_factor} for further details.}

A nuanced aspect of using this asset universe is to repeatedly build the panel each month to handle assets repeatedly entering and leaving the panel. For example, a previously included asset may not be included in January 2019 so we remove the data we have for it, but then the asset could be included in February 2019, for which, we would want to use its historical data (i.e. January 2018-January 2019) to inform our models. Thus, we have to rebuild the panel at a month by month frequency.

There are a few limitations of this approach. We study only a small number of assets relative to the literature. Moreover, we study only assets available to trade for US investors leaving open to future work a more global view of this asset class. Finally, we identify relevant assets at a monthly frequency for simplicity. 

\paragraph{Asset Characteristics} For all assets in the universe, we obtain from Q4 2017 through Q4 2022 a rich set of asset characteristics from several data providers.\footnote{The providers are Coin Metrics, CoinAPI, CoinGecko, CoinMarketCap, Messari, and Santiment.} Tables \ref{t:datadesc_onchain}-\ref{t:datadesc_fin} in Appendix \ref{s:asset_char_desc} enumerate the details of the characteristics studied across six categories: onchain, exchange, social, momentum, microstructure, and financial. We are replicating results, using our more rigorous panel, for many of these characteristics as derived from the relevant literature, including \cite{bianchi2022dynamics}, \cite{borri2022crypto}, \cite{liu2022common}, \cite{liu2021accounting}, \cite{cong2022value}, \cite{liebi2022there}, \cite{zhang2020idiosyncratic}, \cite{zhang2021downside}, \cite{zhang2023liquidity}, and \cite{yao2021investor}. 

To our knowledge, we also have several asset characteristics novel to the literature across the categories: onchain (i.e., age destroyed, delta flow distribution, delta holders distribution, percentage of supply in profit), exchange (i.e., percentage of supply on various exchanges and exchange flows), social (i.e., positive and negative sentiment, developer activity, and VC owned), momentum (i.e., returns from all time highs and lows), microstructure (i.e., ask and bid sizes), and financial (i.e., market capitalization to realized value).

To handle missing values in the characteristics, we fill with cross-sectional medians, where a characteristic was dropped if it had a week where the majority of assets were missing a value. Although this is restrictive, our aim was to be conservative.

\paragraph{Panel Statistics} To close the description of the data, we present basic statistics for our panel of crypto asset excess returns. Summary statistics are presented in Table \ref{f:summary_stats} for the panel's asset returns, market capitalizations, and trading volume. In Panel A, we see the number of assets in the panel begins at 10 in January 2018 and grows to 204 at the end of 2022 with a total of 210 unique assets in the panel. The total market capitalization of our panel captures above 80\% of the total crypto market capitalization as reported by CoinMarketCap. The median weekly asset trading volume is in the tens of millions of USD across the study time period. In Panel B, we report annualized excess return statistics for the market-weighted return of the crypto assets in the panel (CMKT) at 53.84\% per year and a Sharpe ratio of 0.67, which offers a higher absolute and risk-adjusted return than the Nasdaq at 9.85\% and 0.43, respectively.

\begin{figure}[p]
\caption{Empirical distributions of CMKT, Bitcoin, and Ethereum weekly returns.}
\includegraphics[width=\textwidth]{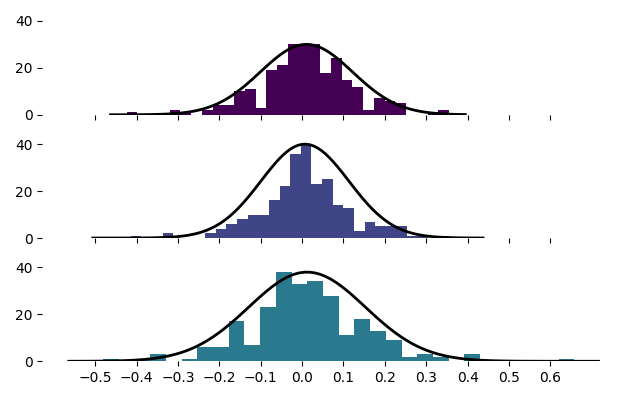}
\note{This figure shows the empirical distributions of weekly excess returns, with a normal distribution fit, for coin market (top panel), Bitcoin (middle panel), and Ethereum (bottom panel) for the January 1, 2018 to December 31, 2022 period.}
\label{f:emp_dists}\end{figure}

Figure \ref{f:emp_dists} reports the empirical distributions of weekly excess returns for the crypto market, bitcoin, and ethereum. The distributions have positive mean with the vast majority of returns between -30\% and 30\%. Across all three, we see several outlier weekly returns that would be unlikely under a normal DGP. Bitcoin has a tighter distribution than the CMKT. Finally, Ethereum interestingly has a right tailed skewness.

\begin{figure}[p]
\caption{Cumulative weekly return of assets in universe.}
\includegraphics[width=\textwidth]{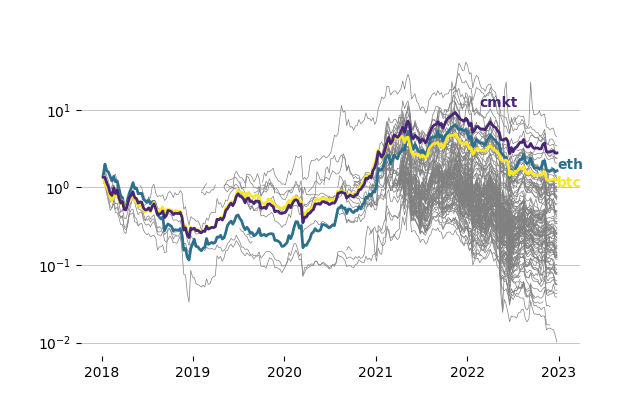}
\note{This figure shows the cumulative excess returns for each asset in the study's universe, and the crypto market, for the January 1, 2018 to December 31, 2022 time period.}
\label{f:cum_returns}\end{figure}

Figure \ref{f:cum_returns} reports the cumulative excess returns from January 1, 2018 until December 31, 2022 for each asset in the study universe and the crypto market. We thus see 172 of the 210 unique assets ($\sim$82\%) have a negative cumulative return over the study period, while 146 and 57 assets ($\sim$70\% and $\sim$27\%) had a cumulative return below -50\% and -90\%, respectively. 

Finally, Tables \ref{f:char_desc_stat_panela} and \ref{f:char_desc_stat_panelb} report descriptive statistics for the panel's dependent variable, asset excess returns over the subsequent week, and the set of sixty three asset characteristics.

\section{Motivating Empirical Facts}\label{s:empfacts}

\begin{quotation}
``We have always had bad money because private enterprise was not permitted to give us a better one...The important truth to keep in mind is that we cannot count on intelligence or understanding but only on sheer self-interest to give us the institutions we need.'' 

---Friedrich A. \citet{hayek1976denationalization} \textit{The Denationalization of Money}.
\end{quotation}

\begin{figure}[h]
\caption{Market Caps (USD).}
\includegraphics[width=\textwidth]{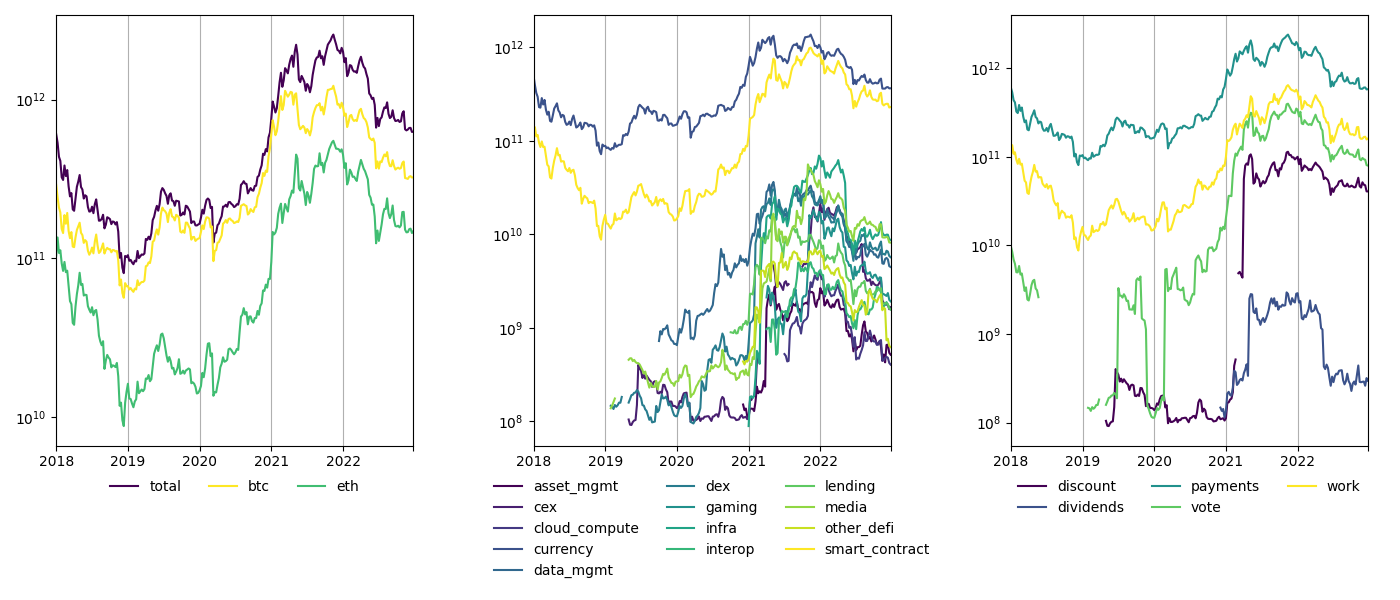}
\note{This figure shows the market capitalization of: in the first panel, the entire panel, Bitcoin, and Ethereum; in the second panel, the entire panel by asset industry classification; and, in the third panel, the entire panel by asset usage classification. The asset industry and usage classifications are from \href{https://messari.io/}{Messari}.}
\label{f:mcaps}\end{figure}

Tables and Figures \ref{f:mcaps}-\ref{f:mi_panelb} present the following eleven motivating empirical facts.
\begin{enumerate}
    \item From zero in 2009, Bitcoin and hundreds of other crypto assets have become a trillion dollar asset class in 2022, with several multi-billion dollar sub-industries.
    \item Bitcoin achieved superior risk-adjusted returns for nearly the entire study time period as compared to traditional asset classes.
    \item Bitcoin has lower correlations to the Nasdaq and S\&P500 (at 0.23 and 0.21) than that of gold's correlation to these indices (at 0.26 and 0.28).
    \item Bitcoin's correlation with other assets is highly time varying, including several quarters of zero or negative correlation with the Nasdaq; their high correlation (> 0.3) is only observed recently in 2022.
    \item From diversifying a risk portfolio of holding 100\% Nasdaq to 60\% Nasdaq and 40\% CMKT, one would obtain a Sharpe Ratio gain of 0.53 (from 0.43 to 0.96).
    \item The crypto market offers a positive inflation risk premium of 31 bps.
    \item Bitcoin is used to store value by a majority of wallets, not speculatively trading.
    \item Bitcoin is a payment network settling hundreds of billions of dollars annually where the large majority of transactions cost less then one USD.
    \item Efforts to fork, that is copy, the Bitcoin blockchain have had immaterial adverse effects on it; an event study of forks observes, on the contrary, significant positive effects on price, trading volume, active addresses, and social activity.
    \item There are several characteristics with significant signal for the cross-section of one-week ahead expected returns.
    \item The asset characteristics contain redundant information; however, the variation cannot be captured by just a few principal components.
\end{enumerate}

\paragraph{A Rising Asset Class} We begin with documenting the birth and rise of crypto asset class. Figure \ref{f:mcaps} plots the market capitalization, during the study period, of the crypto market, Bitcoin, Ethereum, assets by industry classification, and assets by usage classification. Before the launch of Bitcoin, there were, on the order of, 50 global currencies \citep{fratzscher2009explains}. From a market value of zero in 2009, Bitcoin and a Cambrian explosion of hundreds of other cryptocurrencies and crypto assets have risen to a trillion dollar asset class at the end of 2022, with several multi billion dollar sub-industries. The assets included in this study have an aggregate valuation of about \$650B as of December 25, 2022, which captures about 80\% of the total crypto market, per CoinGecko, as of that date. Bitcoin and Ethereum capture over half of the valuation of this asset class, hence the focus on these assets throughout this paper. In the second two panels of Figure \ref{f:mcaps}, we document, using the asset industry and usage classifications of Messari, the rise of several different sub-industries within crypto. We note this not only to emphasize how it is incorrect to conceptualize all crypto assets as cryptocurrencies, but also to note areas for future work to account for these industry and usage differences in asset pricing models and other research.

\begin{figure}[p]
\caption{Sharpe Ratios: Bitcoin vs Major Asset Classes.}
\includegraphics[width=5in]{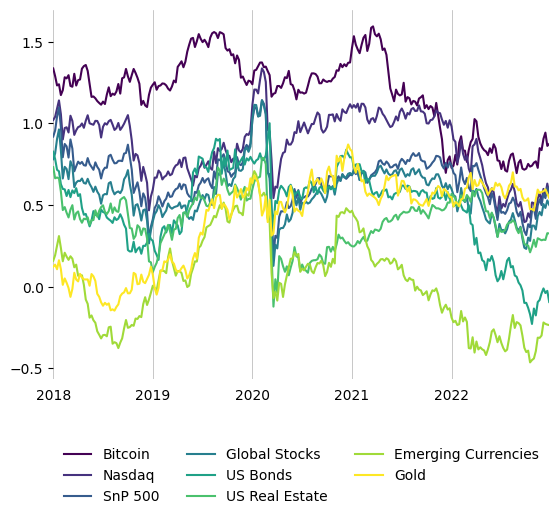}
\note{This figure shows the rolling Sharpe Ratio over four year trailing windows using weekly excess returns for various asset classes. Bitcoin is the weekly return from Kraken's order book. Nasdaq and SnP 500 are the returns of the respective indices. The remaining series correspond to the following ETFs: Global Stocks is VT; US Bonds is BND; US Real Estate is VNQ; Emerging Currencies is EBND; and, Gold is GLD.}
\label{f:sharpe}\end{figure}

\paragraph{An Attractive Asset Class} Bitcoin achieved superior risk-adjusted returns for nearly the entire study time period as compared to traditional asset classes. Figure \ref{f:sharpe} reports rolling Sharpe ratios over trailing four year windows using weekly excess returns for bitcoin and various other assets for the study period. To proxy for other assets (i.e., equities, fixed income, real estate, currencies, and gold), various large ETFs were used to capture a low cost exposure to the relevant asset class. Qualitatively, these rolling Sharpe ratios increase in noise as we shrink the window over which we consider an investment horizon; notably, Bitcoin's separation decreases. However, we study this four year window as it is the relevant period over which the median bitcoin is held, as will be discussed shortly. 

\paragraph{An Independent Asset Class} We now motivate the independence of this asset class by studying correlations and the gains from diversification. Table \ref{f:corr_table} reports pairwise Pearson correlation coefficients between weekly excess returns of bitcoin, ethereum, the crypto market, and various other assets for the January 1, 2018 to December 31, 2022 time period. Bitcoin has lower correlations to the Nasdaq and S\&P500 (at 0.23 and 0.21) than that of gold's correlation to these indices (at 0.26 and 0.28). The crypto market is of similar correlations to the Nasdaq and S\&P500 (at 0.26 and 0.25) as Gold. This is suggestive evidence of the independence of this asset class; however, this single measure masks significant temporal variation in this relationship between assets.

\begin{figure}[p]
\caption{Rolling Four Year Pearson Correlations: Bitcoin vs Major Asset Classes.}
\includegraphics[width=5in]{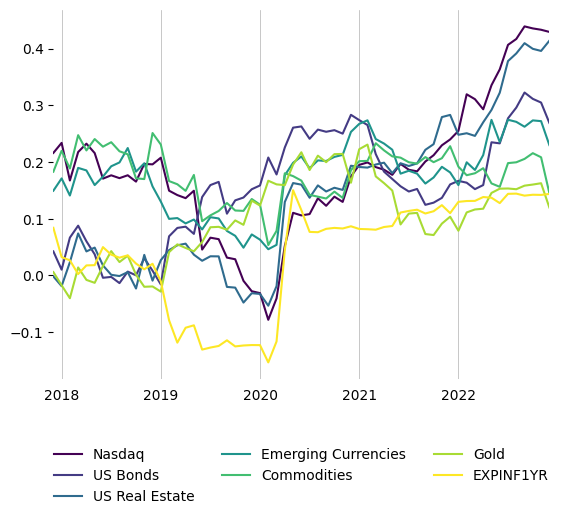}
\note{This figure shows rolling four-year Pearson Correlation coefficients between Bitcoin's weekly excess returns and those of other major asset classes for the January 1, 2018 to December 31, 2022 time period. Nasdaq refers to the weekly excess return of the IXIC index. US Bonds, US Real Estate, Emerging Currencies, Commodities, and Gold refer to the weekly excess returns of the following ETFs: BND, VNQ, EBND, DBC, and GLD. EXPINF1YR refers to the Federal Reserve's measure of expected inflation over the subsequent year.}
\label{f:corr_figure}\end{figure}

Table \ref{f:corr_figure} reports rolling four-year Pearson Correlations between bitcoin's weekly excess returns and those of other major assets for the study time period. We observe those aggregate positive correlations mask richer temporal variation in this rolling correlation. For example, for twelve months leading up to the COVID 19 onset, Bitcoin had a rolling four-year correlation with Nasdaq of less than 0.1, including a quarter of negative correlation. Correlations between Bitcoin and other assets above 0.3 are only a recent phenomenon in the 2022 calendar year. With risk-adjusted returns and correlation measures as context, we now turn to study a better measure of the independence of this asset class: are there gains from diversifying one's portfolio to include crypto?

\begin{figure}[p]
\caption{Crypto Asset's Annualized Cumulative Returns and Volatility.}
\includegraphics[width=\textwidth]{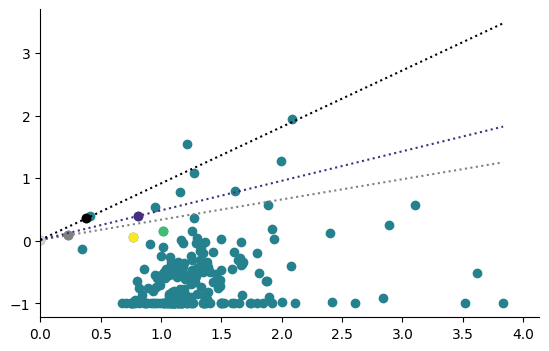}
\note{This figure shows the annualized cumulative return and annualized volatility of simple weekly excess returns of the crypto assets in the study universe (over the 2018-2022, inclusive, time period) as well as a few other portfolios. The light grey point is the risk free rate captured by the annualized cumulative return of the 1 month treasury bill during the study period. The grey point is the annualized cumulative return and annualized volatility of the Nasdaq index. For the same two measures, the yellow point corresponds to BTC, the light green point for ETH, and the purple point for CMKT. For the same two measures, the black point corresponds to a portfolio holding 60\% Nasdaq and 40\% CMKT. The remaining dark green points are for the rest of the assets in the study, removing three assets with outlier returns: DOGE at 9x, LUNA at 52x, and MATIC at 18x.}
\label{f:risk_return}\end{figure}

Figure \ref{f:risk_return} plots the annualized geometric average return against the annualized volatility of each crypto asset's weekly excess returns over the study period.\footnote{Do note the ordinate axis is a geometric average, not a time series average, given the latter can be quite different from the former given the volatility of these assets (e.g. Zcash has different signs!).} Additionally, the risk free rate and the portfolio holding 100\% Nasdaq are plotted. Nasdaq, in dark grey, and CMKT, in purple show similar ratios between their annualized geometric average returns and risk with the grey and purple dashed lines stacking nearly on top of each other. However, if we aim to maximize this return-risk ratio in this data set, we should allocate 70\% to the Nasdaq and 30\% to the crypto market for a ratio of 0.63, which is the portfolio plotted in black. (For the Sharpe Ratio, we'd maximize at 60\% Nasdaq and 40\% CMKT for a Sharpe of 0.96, a Sharpe gain of 0.53 over 100\% Nasdaq at 0.43.) In this data set, the annualized geometric average weekly return of CMKT is 26.4\% with an annualized volatility of 80.8\%. For BTC, ETH, Nasdaq, and the 70 30 portfolio, those numbers are, respectively: 5.9\% and 76.7\%; 11.8\% and 101.8\%; 7.5\% and 23.0\%; and, 20.5\% and 32.4\%.

\paragraph{Inflation Risk Premium in the Crypto Asset Class} Allocations to bitcoin and the crypto market have been motivated as a hedge on traditional currencies; for example, inflation risk as substantiated by recent survey data \citep{aiello2023invests}. Some simple methods to empirically study this claim would be to measure correlations between crypto returns and inflation expectations or, similarly, a price impact study of large changes in inflation expectations on crypto returns. In the spirit of these, if we study only the largest twelve month changes in the Cleveland Fed's 10 year expected inflation measure between January 1 2018 and December 31, 2022, the correlation between bitcoin monthly excess returns and 10 year expected inflation is 0.03 and the correlation between cmkt monthly excess returns and 10 year expected inflation is 0.06, which compared to monthly excess return of Gold and 10 year expected inflation is -0.10. The next level of sophistication would be to partition out the crypto market return in a regression of the monthly excess returns of Bitcoin on the CMKT and expected inflation, which is reported in Panel A of Table \ref{f:inflation}. We observe a positive covariance between inflation expectations and the contemporaneous excess returns of bitcoin. However, in asset pricing, we have a better tool to capture whether an observable risk factor, e.g., inflation, carries a nonzero risk premium in the crypto asset class.

In Table \ref{f:inflation}, Panel B, we report the positive inflation risk premium of 31 bps per week for the crypto asset class. We utilize Fama Macbeth regressions, which recover the inflation-mimicking portfolio in the crypto asset class and estimate this portfolio's risk premium. Although this yields a positive and economically significant risk premium, it is not statistically distinguishable from zero. Other researchers however have found a similarly positive premium associated with U.S. inflation breakeven rate \citep{borri2022crypto}. A limitations of note is that we used a 200 day restriction to calculate the factor loading \citep{bali2016empirical}.

\paragraph{Onchain Facts on Bitcoin's Use} We now turn to study a few onchain statistics, using the rich environment of the Bitcoin blockchain, with the aim to correct a few misconceptions in the literature and to motivate future research in the onchain lab.

\begin{figure}[h]
\caption{Hodling: Bitcoin UTXO Median Age in Days.}
\includegraphics[width=3.7in]{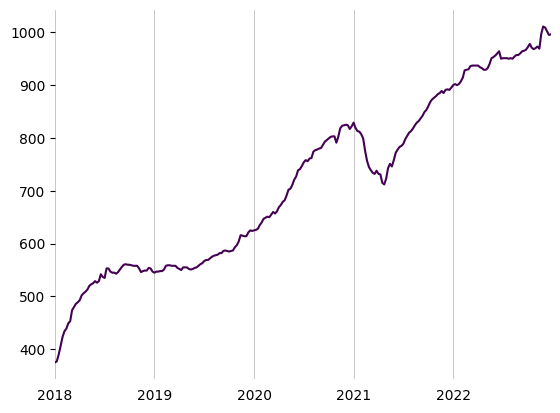}
\note{This figure shows median age in full days of all unspent transaction outputs (UTXOs), rounded down to the nearest day, on the Bitcoin ledger for each week in 2018 to 2022, inclusive. \href{https://bitcointalk.org/index.php?topic=375643.0}{Why hodl?}}
\label{f:hodling}\end{figure}

First, Bitcoin is used to store value by a majority of wallets, rather than speculatively trading. Figure \ref{f:hodling} reports the median age in full days of all unspent transaction outputs (UTXO) for the Bitcoin ledger for each week in 2018 through 2022. The majority of UTXOs, which are a proxy for wallets, have not been spent for years. This may be a downward biased estimate given Bitcoin has only been around for about fifteen years, which the trend line supports. 

\begin{figure}[h]
\caption{Bitcoin Onchain Transactions.}
\includegraphics[width=3.7in]{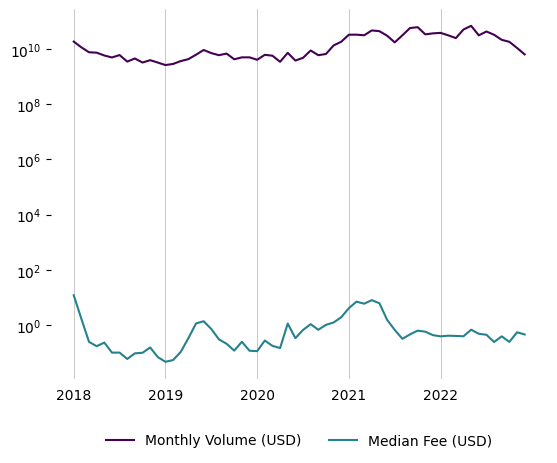}
\note{This figure shows two time series for onchain bitcoin transactions. Monthly Volume reports, in USD, the total calendar month onchain volume transferred between distinct addresses. The Median Fee reports, in USD, the median fee paid to miners across all transactions within each calendar month.}
\label{f:btc_tx}\end{figure}

Second, Bitcoin is a payment network settling hundreds of billions of dollars annually where the large majority of transactions cost less then one USD, offering monetary functions with distributed consensus. Figure \ref{f:btc_tx} reports Bitcoin's monthly onchain volume and median transaction fee for the study period. We see monthly onchain settled transactions on the order of tens of billions of USD and median transaction fees paid to miners on the order of one USD.

Third, efforts to fork, that is copy, the Bitcoin blockchain have had immaterial adverse effects on it. Table \ref{f:forks} reports this an event study for various Bitcoin statistics on fifteen dates of major Bitcoin forks, subsequent to January 2016.\footnote{We use one week pre and post windows. The fifteen forks considered are: Bitcoin 21, Zcash, Bitcoin Cash, Bitcoin Gold, Bitcoin Diamond, Bitcoin Lightning, Bitcoin Fast, Bitcoin2, BitcoinPlus, Bitcoin Interest, Bitcoin Atom, Bitcoin Private, Microbitcoin, Bitcoin BEP2, and BitcoinSV. These are major assets who had a market price tracked by CMC at some point in time subsequent to the fork.} Bitcoin's market value, trading volume, onchain and development activity, social volume, and miner hash rate all respond positively to forks with statistically significant positive signs for return, trading volume, active addresses, and social volume. \footnote{There are a few major Bitcoin forks not included (e.g. Litecoin), which were outside the time period where we have non-financial data.} \footnote{To check robustness, we also ran the event study for two day periods before and after event days as well as two week periods before and after. The only qualitative difference in a point estimate were Return and Miner Hash Rate flipping negative for two week windows, albeit both are not statistically indistinguishable from zero. All other statistics maintain sign and significance.}

\paragraph{Asset Characteristic Signal Content} Finally, we study some simple statistics to inform the potential signal content of the sixty three asset characteristics in measuring expected returns. Tables \ref{f:onchain}-\ref{f:mi_panelb} report statistics for the panel's asset characteristics, including: univariate regression results of asset excess returns over several forward horizons on each characteristic; the mutual information between asset excess returns seven days ahead and each characteristic for entire study period and each calendar year; and, the correlations among various groups of characteristics and their first principal components. Although decaying over the years, we observe numerous sources of signal for the cross-section of one-week ahead expected returns. Further, the asset characteristics do contain redundant information; however, the variation across asset characteristics cannot be captured by just a few principal components.

Across the univariate regression results in tables in \ref{f:onchain} through \ref{f:financial_panelb}, we observe numerous significant coefficients not only contemporaneously and at the horizon to be studied (i.e. seven days ahead), but also that the panel exhibits longer memory with significant results at 14, 30, and 90 day horizons. That is, there is persistent signal in characteristics for returns over multiple future horizons beyond seven days. There are more significant coefficients than we would expect by chance and, moreover, there are numerous coefficients that remain highly significant at all horizons.

In the tables \ref{f:mi_panela} and \ref{f:mi_panelb}, we study mutual information to capture a broader, nonlinear measure of the relationship between the asset characteristics and the subsequent seven day asset excess return. Correlations and univariate regression coefficients capture only a simple linear covariance. In these tables, we see similar patterns to the univariate regression results for characteristics with high signal; however, we also observe falling magnitudes in the mutual information.

Across the correlation tables in \ref{f:onchain} through \ref{f:financial_panela}, we observe the redundancy in information with many pairs of characteristics having near one Pearson correlation; in the \ref{f:corr_pc}, we see the first principal components for each asset characteristic category has some redundancy, however the majority of the pairwise relationships show low correlation; and, finally, in the characteristic correlation tables, the first principal components can capture large variation only among a few of the characteristics.

\section{Univariate Observable Factor Models}\label{s:uni_factor}

We next study the cross-section of returns with a classic nonparametric method of univariate factor portfolio sorts based on the rich set of crypto asset characteristics. We establish a set of empirical patterns in the dynamic of crypto returns to offer suggestive evidence of why different crypto assets earn different average returns. Moreover, these results can be used to motivate and develop theoretical models, in addition to more practical and obvious use.

We study portfolios formed using sixty three asset characteristics as precisely defined in Appendix \ref{s:asset_char_desc}. The large majority of these have extensively-studied counterparts in equity markets, and further, as previously discussed in Section \ref{s:descdata}, have a small but growing literature studying these risk factors in the crypto asset class. Nevertheless, to our knowledge, we have at least fifteen novel asset characteristics across the six categories we use to group the sixty three characteristics. They are: age destroyed, delta flow distribution, delta holders distribution, percentage of supply in profit, percentage of supply on various exchanges, exchange flows, positive and negative sentiment, developer activity, VC owned, returns from all time highs and lows, ask and bid sizes, and market capitalization to realized value.

To form univariate factor portfolios for the study period 2018-2022, inclusive, we sort crypto assets each week into five value-weighted portfolios ranked by the smallest (portfolio 1) values for the characteristic to the largest (portfolio 5). We analyze the zero-investment portfolio of long the top quintile and short the bottom quintile. That is, each week we sort individual crypto assets into quintile portfolios based on the value of the given characteristic, and then we track the value-weighted excess return of each portfolio over the week.

Table \ref{f:uni_factors_sig} presents the results for the statistically significant zero-investment long-short strategies across all characteristics. Six of sixty three characteristics exhibited statistically significant long-short strategies. Although is a low fraction, this has a low probability of occurring by chance.\footnote{There is a 3.7\% probability of observing 6 or more successes out of 63 independent Bernoulli trials with success probability 0.05\%.} Across the thirty quintile portfolios, twenty four followed a monotonic pattern within characteristic, with only the two week momentum having monotonic quintile portfolios. The time-series average weekly excess return spread (and annualized Sharpe ratios) for the zero-investment long-short strategies are 1.5\% (0.78) for two week momentum, 1.2\% (0.75) for one month industry momentum, 1.4\% (0.87) for two month industry momentum, 1.6\% (0.76) for one week beta, 1.2\% (0.79) for one month idiosyncratic skewness, and 1.4\% (0.76) for one week 5\% expected shortfall.

It is of note that all of these statistically significant strategies were formed on functions of previous returns, that is, momentum and financial characteristics. Although capturing differences in industries lead to significant long momentum strategies, more fundamental pricing characteristics were not associated with significant spreads between top and bottom quintile portfolios. With this more rigorous panel, many of the univariate factor results established in the existing literature fail to replicate. Many characteristics have economically significant return spreads and Sharpe ratios, although the short panel of five years and the volatility of returns in this asset class suppress the resulting \textit{t}-statistics.

Tables \ref{f:uni_factors_onchain}-\ref{f:uni_factors_fin} present the results for the statistically insignificant zero-investment long-short strategies across the remaining characteristics, grouped into separate tables by characteristic category. Five more of these characteristics had an economically meaningful annualized Sharpe ratio above 0.5 for the associated zero-investment long-short strategy. For these, the time-series average weekly excess return spread (and annualized Sharpe ratios) for the zero-investment long-short strategies are 0.8\% (0.61) for age destroyed, -1.0\% (0.58) for delta holders distribution, -1.0\% (0.57) for three to two month reversal, 1.0\% (0.56) for one month alpha, and -1.1\% (0.54) for one month beta. Seventeen more of the characteristics had an economically meaningful annualized excess return above 30\% for the zero-investment long-short strategy.\footnote{These are pure-alpha strategies, although for context the CMKT and Nasdaq returned 0.53\% and 10\% annually during the same time period.} For these, the time-series average weekly excess return spread for the zero-investment long-short strategies are 0.6\% for one week transaction volume, 0.7\% for percent of circulating supply on a CEX, 0.7\% for percent of circulating supply on DeFi, 0.7\% for Reddit social volume, 0.8\% for one week momentum, 0.6\% for one month momentum, -0.6\% for return from all time high, -1.0\% for return from all time low, 0.9\% for one week alpha, 0.7\% for one month downside beta, 0.7\% for one month coskewness, 0.9\% for one week Var5\%, -0.6\% for one week volatility, -0.6\% for one month volatility, -0.9\% for three month volatility, -0.5\% for one week idiosyncratic volatility, and -0.6\% for one month idiosyncratic volatility.

We observe several notable patterns. There is a large Sharpe of economically significant return spreads and Sharpe ratios for the long-short strategies formed using financial characteristics, yet none for microstructure characteristics. Although using these more relaxed economically significant thresholds, there were several onchain characteristics of note, which is promising for more fundamental-based asset pricing. Only a single social characteristic had an economically significant return spread, with no stronger results.

We close this section noting limitations. These univariate factor results do not account for feasibility of short selling and transaction costs, such as spreads, trading fees, margin fees, price impact, and slippage. We also studied results for the entire study period, without an formal out of sample exercise.

\section{Conclusion}\label{s:conclusion}

\begin{quotation}

``Competition would provide better money than would government. I believe we can do much better than gold ever made possible. Free enterprise, i.e. the institutions that would emerge from a process of competition in providing good money, no doubt would.

Two hundred years ago in \textit{The Wealth of Nations} Adam Smith wrote that

`to expect, indeed, that the freedom of trade should ever be entirely restored in Great Britain, is as absurd as to expect that an Oceana or Utopia should ever be established in it.'

It took nearly 90 years from the publication of his work in 1776 until Great Britain became the first country to establish complete free trade in 1860...I fear that since `Keynesian' propaganda has filtered through to the masses, has made inflation respectable and provided agitators with arguments which the professional politicians are unable to refute, the only way to avoid being driven by continuing inflation into a controlled and directed economy, and therefore ultimately in order to save civilisation, will be to deprive governments of their power over the supply of money. What we need now is a Free Money Movement...

I wish I could advise that we proceed slowly. But the time may be short. What is now urgently required is not the construction of a new system but the prompt removal of all legal obstacles which have for two thousand years blocked the way for an evolution which is bound to throw up beneficial results which we cannot now forsee.''

---Friedrich A. \citet{hayek1976denationalization} \textit{The Denationalization of Money}.
\end{quotation}

\newpage
\section{References} 
\bibliography{\bib}

\newpage
\appendix

\section{Details on Crypto Asset Characteristics}\label{s:asset_char_desc}

The tables below present details for each category of crypto asset characteristics.

\begin{table}[h!]
\scriptsize
\caption{Crypto Asset Onchain Characteristics}
\begin{tabular}{ l p{0.7\textwidth} } 
\toprule
Characteristic & Definition \\
\midrule
Tx Volume Tm7 & The total transaction volume in native units over the trailing seven days. \\
Active Addresses Tm7 & The number of active address over the trailing seven days. \\
$\Delta$ Log New Addresses Tm14-Tm7 & The first difference of the logarithm of new addresses from 14 to 7 days ago.\\
New Addresses Tm7 & The total number of new addresses over the trailing seven days. \\
Total Addresses & The total number of unique addresses.\\
Circulation Tm7 & The number of unique native units transferred over the trailing seven days.\\
Age Destroyed & The sum over the trailing week of all native units transferred times the number of days since they were previously transferred.\\
$\Delta$ Flow Distribution Tm7 & The ratio between the total native units transferred between various entities identified by Santiment (e.g. cex, dexes, defi platforms, whales, etc.) over the trailing week and the total first absolute differences across all the flow variables over the trailing week. \\
$\Delta$ Holders Distribution Tm7 & The same functional form as the change in flow but for the total supply held by wallets with various magnitudes, as identified by Santiment, of the total supply. \\
\% Supply in Profit & The percentage of the total native units which last transferred at a market value below the current market value. \\
\bottomrule
\end{tabular}
\label{t:datadesc_onchain}
\end{table}

\begin{table}[h!]
\scriptsize
\caption{Crypto Asset Exchange Characteristics}
\begin{tabular}{ l p{0.7\textwidth} } 
\toprule
Characteristic & Definition \\
\midrule
\% Circ. Supply CEX & The percentage of circulating supply in native units in wallets associated with CEXs as identified by Santiment. \\
\% Circ. Supply DEX & The percentage of circulating supply in native units in wallets associated with DEXs as identified by Santiment. \\
\% Circ. Supply Defi & The percentage of circulating supply in native units in wallets associated with DeFi platforms as identified by Santiment. \\
\% Circ. Supply Traders & The percentage of circulating supply in native units in wallets associated with active traders as identified by Santiment. \\
Exchange Inflow & The total number of native units transferred from wallets not associated with exchanges to wallets that are, over the trailing week, as identified by Santiment. \\
Exchange Outflow & The total number of native units transferred from wallets associated with exchanges to wallets that are not associated with exchanges, over the trailing week, as identified by Santiment. \\
Number of Trading Pairs & The number of trading pairs identified by CMC on CEXs. \\
\bottomrule
\end{tabular}
\label{t:datadesc_exchange}
\end{table}

\begin{table}[h!]
\scriptsize
\caption{Crypto Asset Social Characteristics}
\begin{tabular}{ l p{0.7\textwidth} } 
\toprule
Characteristic & Definition \\
\midrule
Social Volume & The total number of text documents containing the asset name across Reddit, Twitter, Telegram, and BitcoinTalk  over the trailing seven days. \\
Social Volume Reddit & The total number of text documents containing the asset name on Reddit over the trailing seven days. \\
Social Volume Twitter & The total number of text documents containing the asset name on Twitter over the trailing seven days. \\
Sentiment Pos. Reddit & The total sentiment score across all text documents with a positive sentiment on Reddit over the trailing seven days. \\
Sentiment Pos. Twitter & The total sentiment score across all text documents with a positive sentiment on Twitter over the trailing seven days. \\
Sentiment Neg. Reddit & The total sentiment score across all text documents with a negative sentiment on Reddit over the trailing seven days. \\
Sentiment Neg. Twitter & The total sentiment score across all text documents with a negative sentiment on Twitter over the trailing seven days. \\
Developer Activity & The aggregate number of GitHub actions (e.g. commits, forks, comments, etc.), as identified by CoinGecko, over the trailing seven days. \\
VC Owned & Whether the asset has been funded by a set of prominent venture capitalists as identified by CoinMarketCap. \\
\bottomrule
\end{tabular}
\label{t:datadesc_social}
\end{table}

\begin{table}[h!]
\scriptsize
\caption{Crypto Asset Momentum Characteristics}
\begin{tabular}{ l p{0.7\textwidth} } 
\toprule
Characteristic & Definition \\
\midrule
Return Tm7 & Momentum over the trailing seven days. \\
Return Tm14 & Momentum over the trailing fourteen days. \\
Return Tm30 & Momentum over the trailing thirty days. \\
Return Tm30 & Momentum over the trailing sixty days. \\
Return Tm90 & Momentum over the trailing ninety days. \\
Return Tm14-Tm7 & Short term reversal: difference in return between trailing fourteen and seven days. \\
Return Tm30-Tm14 & Medium term reversal: difference in return between trailing thirty and fourteen days. \\
Return Tm90-Tm30 & Long term reversal: difference in return between trailing ninety and thirty days. \\
Return from ATH & The return since the all time high price. \\
Return from ATL & The return since the all time low price. \\
Return Industry Tm30 & Industry momentum over the trailing thirty days. \\
Return Industry Tm60 & Industry momentum over the trailing sixty days. \\
\bottomrule
\end{tabular}
\label{t:datadesc_mom}
\end{table}

\newpage 

\begin{table}[h!]
\scriptsize
\caption{Crypto Asset Microstructure Characteristics}
\begin{tabular}{ l p{0.7\textwidth} } 
\toprule
Characteristic & Definition \\
\midrule
Trades Sum Tm7 & The total number of CEX trades in the trailing seven days. \\
Volume Sum Tm7 & The dollar CEX trading volume in the trailing seven days. \\
Spread Bps & Spread in basis points. \\
Ask Size & Market value of orders at best ask. \\
Bid Size & Market value of orders at best bid. \\
Illiq Tm7 & The average absolute hourly return over the trailing week divided by the average hourly dollar volume over the trailing week. \\
Turnover Tm7 & The total volume over the trailing week divided by the circulating supply in native units. \\
\bottomrule
\end{tabular}
\label{t:datadesc_micro}
\end{table}

\newpage

\begin{table}[h!]
\scriptsize
\caption{Crypto Asset Financial Characteristics}
\begin{tabular}{ l p{0.7\textwidth} } 
\toprule
Characteristic & Definition \\
\midrule
Price & The market value of one native unit in USD. \\
Size & The market capitalization of all free floating native units in USD. \\
MVRV & The ratio of the market capitalization to the realized value, or the total number of free floating native units times the dollar value at the the time of the last transfer. \\
Alpha Tm7 & Intercept coefficient from regressing hourly excess returns on cmkt hourly returns over the trailing seven days. \\
Alpha Tm30 & Intercept coefficient from regressing hourly excess returns on cmkt hourly returns over the trailing thirty days. \\
Beta Tm7 & Slope coefficient from regressing hourly excess returns on cmkt hourly returns over the trailing seven days. \\
Beta Tm30 & Slope coefficient from regressing hourly excess returns on cmkt hourly returns over the trailing thirty days. \\
Beta Downside Tm30 & Slope coefficient from regressing negative hourly excess returns (or zero) on negative cmkt hourly returns over the trailing thirty days. \\
Coskew Tm30 & The slope coefficient on the squared cmkt term from regressing hourly excess returns on cmkt hourly returns and squared cmkt hourly returns over the trailing thirty days. \\
ISkew Tm30 & The skewness of the residuals from from regressing hourly excess returns on cmkt hourly returns and squared cmkt hourly returns over the trailing thirty days. \\
Shortfall 5\% Tm7 & Average hourly return of the returns below the fifth quantile of the trailing seven day hourly returns. \\
VaR 5\% Tm7 & The fifth quantile of hourly excess returns over the trailing seven days. \\
Vol Tm7 & The standard deviation of hourly excess returns over the trailing seven days. \\
Vol Tm30 & The standard deviation of hourly excess returns over the trailing thirty days. \\
Vol Tm90 & The standard deviation of hourly excess returns over the trailing ninety days. \\
Ivol Tm7 & The standard deviation of the residuals from regressing hourly excess returns on cmkt returns over the trailing seven days. \\
Ivol Tm30 & The standard deviation of the residuals from regressing hourly excess returns on cmkt returns over the trailing thirty days. \\
Ivol Tm90 & The standard deviation of the residuals from regressing hourly excess returns on cmkt returns over the trailing ninety days. \\
\bottomrule
\end{tabular}
\label{t:datadesc_fin}
\end{table}

\newpage
\section{Tables and Figures}\label{s:tables_and_figures}

\begin{table}[h]
\caption{Summary statistics.}
\includegraphics[width=6.2in]{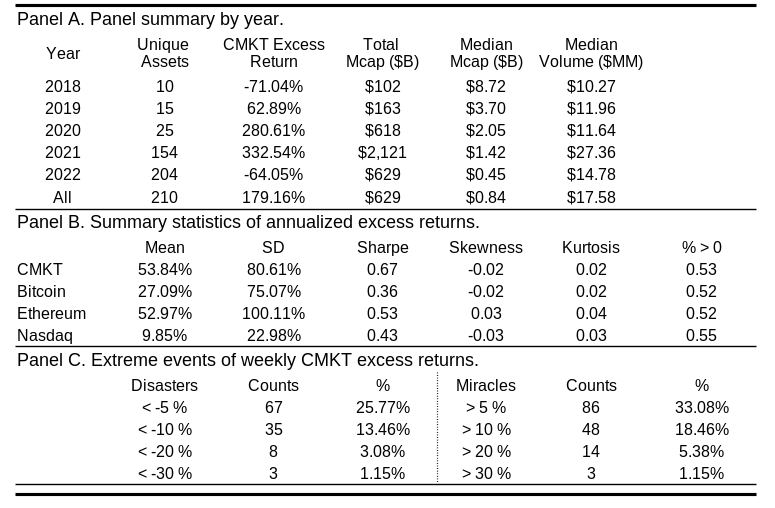}
\note{This table shows summary statistics on the weekly panel of excess returns from January 1, 2018 to December 31 2022. Panel A reports, by calendar year and for the whole panel, the number of unique assets, the cumulative excess return of the crypto market, the total market capitalization in the last week in billions of dollars, the median market capitalization in billions of dollars, and the median weekly volume in millions of dollars. Panel B reports---for the crypto market (CMKT), Bitcoin, Ethereum, and the Nasdaq---annualized excess return statistics, including the mean, standard deviation, Sharpe ratio, skewness, kurtosis, and percentage of weekly excess returns that are positive. Panel C reports the percentage of extreme events using the weekly crypto market index excess returns.}
\label{f:summary_stats}\end{table}

\begin{table}[p]
\caption{Crypto Asset Characteristics: Descriptive Statistics.}
\includegraphics[width=\textwidth]{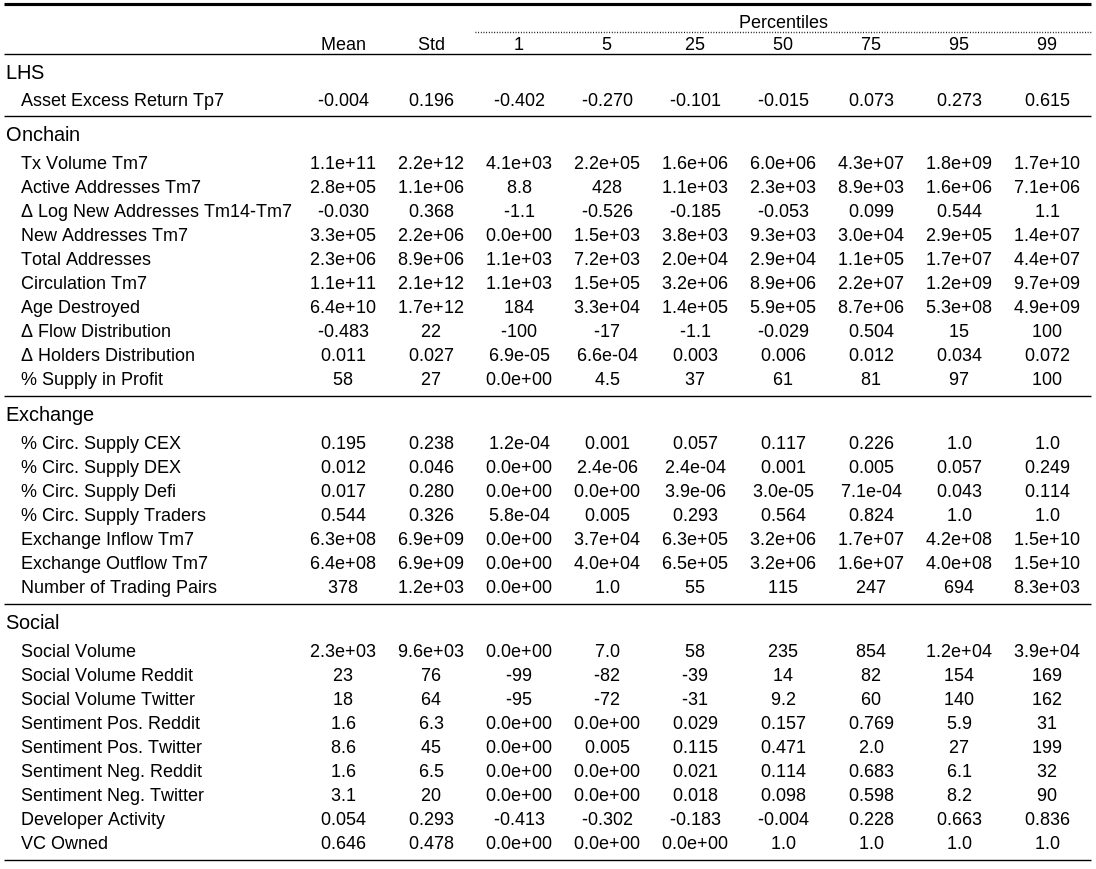}
\note{This table reports the summary statistics from the weekly asset panel for the dependent variable, asset excess returns seven days ahead, and the asset characteristics. For each variable, we report the panel mean, median, standard deviation, and selected percentiles. There are 22,678 asset-weeks from January 7, 2018 to December 15, 2022.}
\label{f:char_desc_stat_panela}\end{table}

\begin{table}[p]
\caption{Crypto Asset Characteristics: Descriptive Statistics (Continued).}
\includegraphics[width=\textwidth]{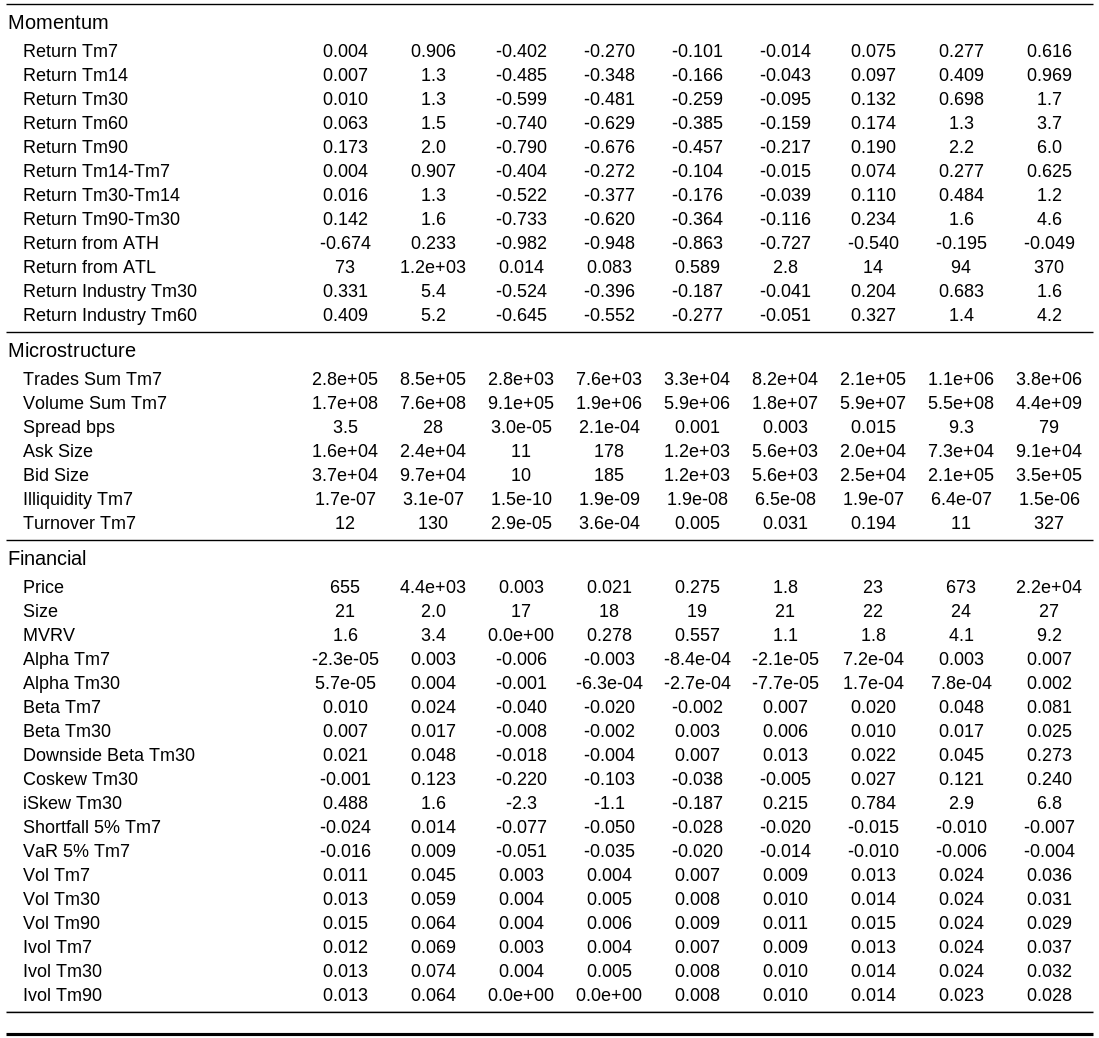}
\note{This table reports the summary statistics from the weekly asset panel for the dependent variable, asset excess returns seven days ahead, and the asset characteristics. For each variable, we report the panel mean, median, standard deviation, and selected percentiles. There are 22,678 asset-weeks from January 7, 2018 to December 15, 2022.}
\label{f:char_desc_stat_panelb}\end{table}

\begin{table}[p]
\caption{Correlations.}
\includegraphics[width=\textwidth]{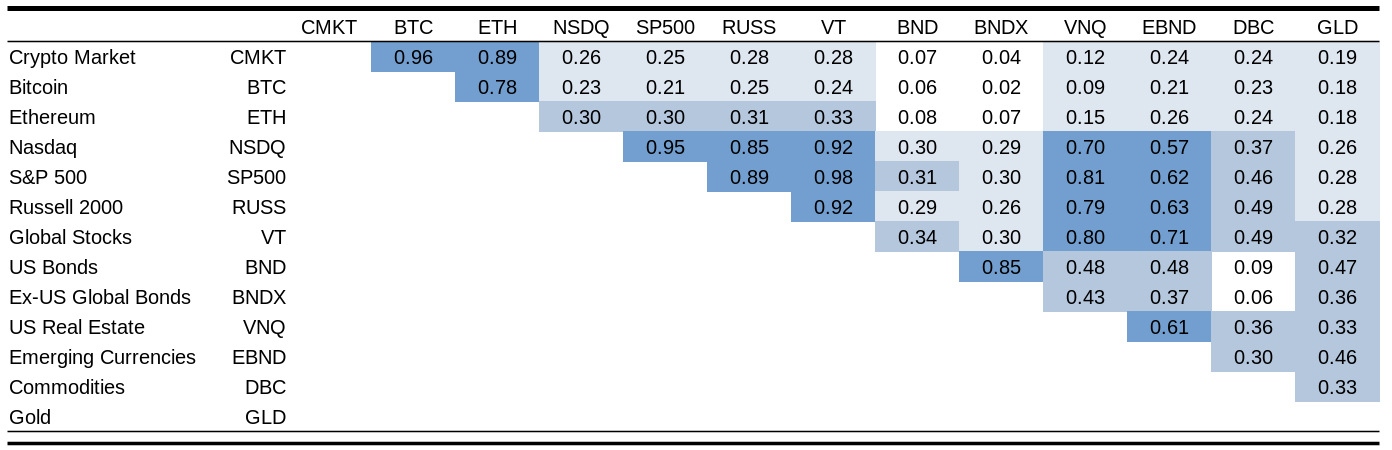}
\note{This table reports Pearson correlation coefficients between weekly excess returns of row and column assets for the January 1, 2018 to December 31, 2022 time period. CMKT, Bitcoin, and Ethereum refer to the weekly excess returns of the market cap-weighted assets in the study universe, bitcoin, and ether, respectively. Nasdaq, S\&P500, and Russel 2000 refer to the weekly excess return of the IXIC, GSPC, and RUT indices, respectively. Global Stocks, US Bonds, Ex-US Global Bonds, US Real Estate, Emerging Currencies, Commodities, and Gold refer to the weekly excess returns of the following ETFs: VT, BND, BNDX, VNQ, EBND, DBC, and GLD.}
\label{f:corr_table}\end{table}

\begin{table}[h]
\caption{Inflation Risk Premium.}
\includegraphics[width=3.5in]{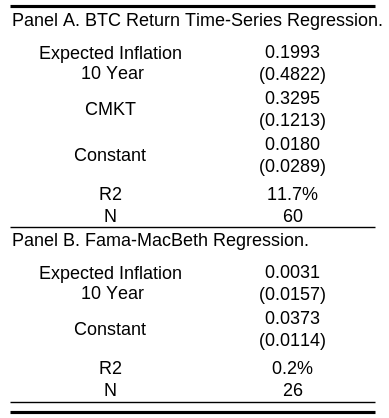}
\note{This table reports results from two regressions. Panel A reports point estimates and standard errors from the time-series regression of BTC monthly excess returns on 1 year expected inflation innovations, CMKT monthly excess returns, and a constant. Panel B reports the results from a Fama-MacBeth regression procedure to estimate the risk premium of inflation in the crypto asset class, where we use assets with at least two years of data to precisely estimate beta hats.}
\label{f:inflation}\end{table}

\begin{table}[h]
\caption{Bitcoin Forks: Event Study.}
\includegraphics[width=5in]{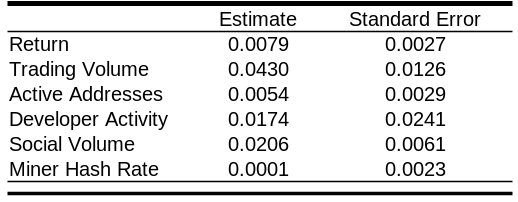}
\note{This table reports an event study for various Bitcoin statistics on dates on which fifteen major Bitcoin forks occurred, subsequent to January 2016. The point estimates are the difference between, in the seven days before and after the event date, the average daily change of each characteristic. Return is the daily change in bitcoin's USD price. Trading Volume is the daily change in bitcoin trading volume reported as by CoinMarketCap. Active Addresses is the daily change in the number of unique active Bitcoin addresses as reported by Santiment. Developer Activity is the daily change in the total number of GitHub events (e.g. code pushes, issue interactions, pull requests, comments on commits, etc.) as reported by Santiment. Social Volume is the daily change in the total number of text documents across Reddit, Telegram, Twitter, and BitcoinTalk containing the keyword ``bitcoin'' as reported by Santiment. Miner Hash Rate is the daily change in the total Bitcoin hash rate as imputed by Coinmetrics. Standard errors are bootstrapped: the standard deviation of the distribution formed by calculating each statistic for 10,000 randomly sampled, with replacement, event days.}
\label{f:forks}\end{table}

\begin{table}[p]
\caption{Onchain Characteristics: Correlations and Signal.}
\includegraphics[width=5.6in]{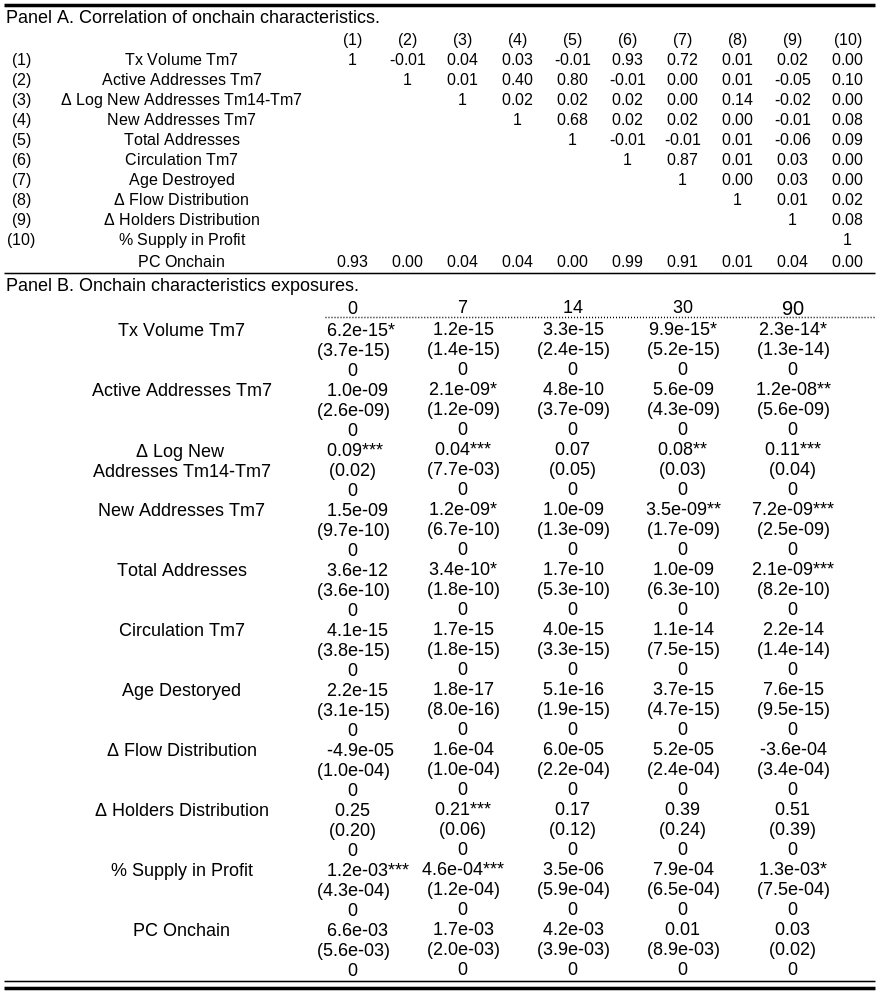}
\note{{\scriptsize This table reports the correlation matrix among Onchain Characteristics and the loadings on asset excess returns on each characteristic at various horizons. Panel A reports pairwise Pearson correlation coefficients among the characteristics and the first principal component from them. The characteristics are re-scaled to be mean zero and unit variance before PCA and studying these correlations. Panel B reports the coefficient (with 1, 2, and 3 stars for significant at the 10\%, 5\%, and 1\% levels, respectively), standard error, and $R^2$ for univariate panel regressions of asset excess weekly returns at 0, 7, 14, 30, and 90 days ahead on each of the characteristics and a constant. Standard errors are Newey-West adjusted using Bartlett's formula for the number of lags. There are 22,678 observations.}}
\label{f:onchain}\end{table}

\begin{table}[p]
\caption{Exchange Characteristics: Correlations and Signal}
\includegraphics[width=\textwidth]{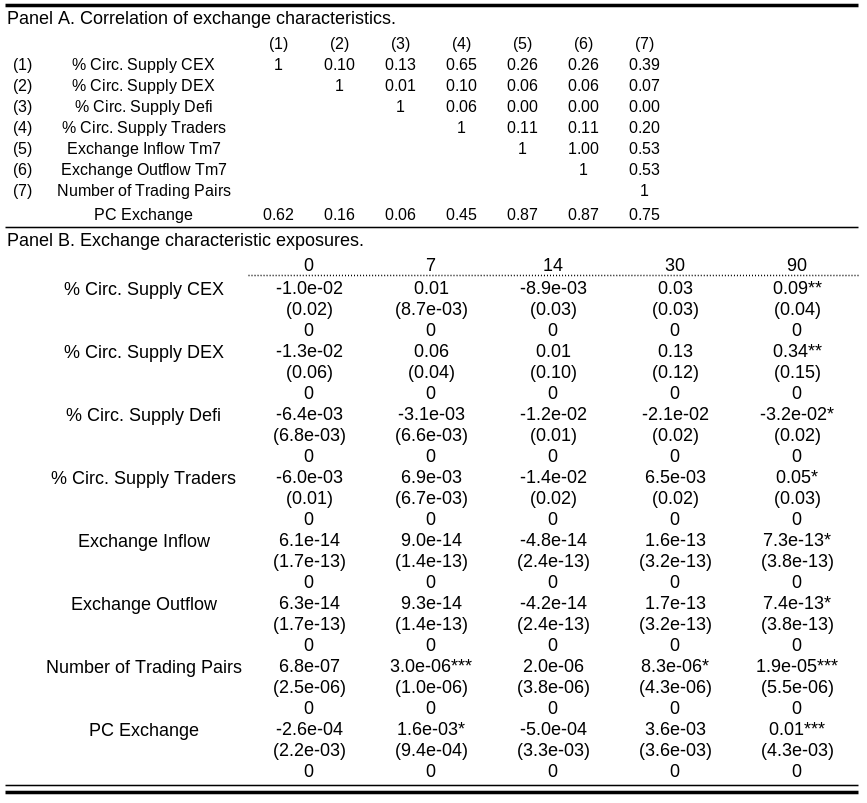}
\note{This table reports the correlation matrix among Exchange Characteristics and the loadings on asset excess returns on each characteristic at various horizons. Panel A reports pairwise Pearson correlation coefficients among the characteristics and the first principal component from them. The characteristics are re-scaled to be mean zero and unit variance before PCA and studying these correlations. Panel B reports the coefficient (with 1, 2, and 3 stars for significant at the 10\%, 5\%, and 1\% levels, respectively), standard error, and $R^2$ for univariate panel regressions of asset excess weekly returns at 0, 7, 14, 30, and 90 days ahead on each of the characteristics and a constant. Standard errors are Newey-West adjusted using Bartlett's formula for the number of lags. There are 22,678 observations.}
\label{f:exchange}\end{table}

\begin{table}[p]
\caption{Social Characteristics: Correlations and Signal.}
\includegraphics[width=6.2in]{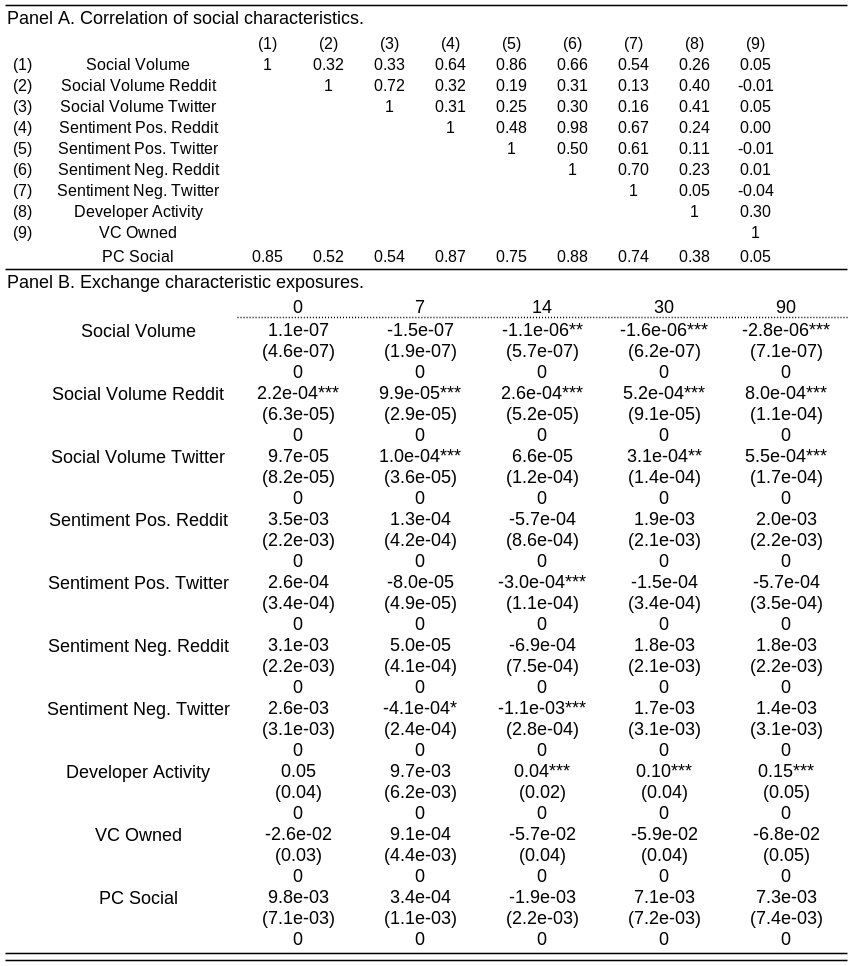}
\note{This table reports the correlation matrix among Social Characteristics and the loadings on asset excess returns on each characteristic at various horizons. Panel A reports pairwise Pearson correlation coefficients among the characteristics and the first principal component from them. The characteristics are re-scaled to be mean zero and unit variance before PCA and studying these correlations. Panel B reports the coefficient (with 1, 2, and 3 stars for significant at the 10\%, 5\%, and 1\% levels, respectively), standard error, and $R^2$ for univariate panel regressions of asset excess weekly returns at 0, 7, 14, 30, and 90 days ahead on each of the characteristics and a constant. Standard errors are Newey-West adjusted using Bartlett's formula for the number of lags. There are 22,678 observations.}
\label{f:social}\end{table}

\begin{table}[p]
\caption{Momentum Characteristics: Correlations and Signal.}
\includegraphics[width=\textwidth]{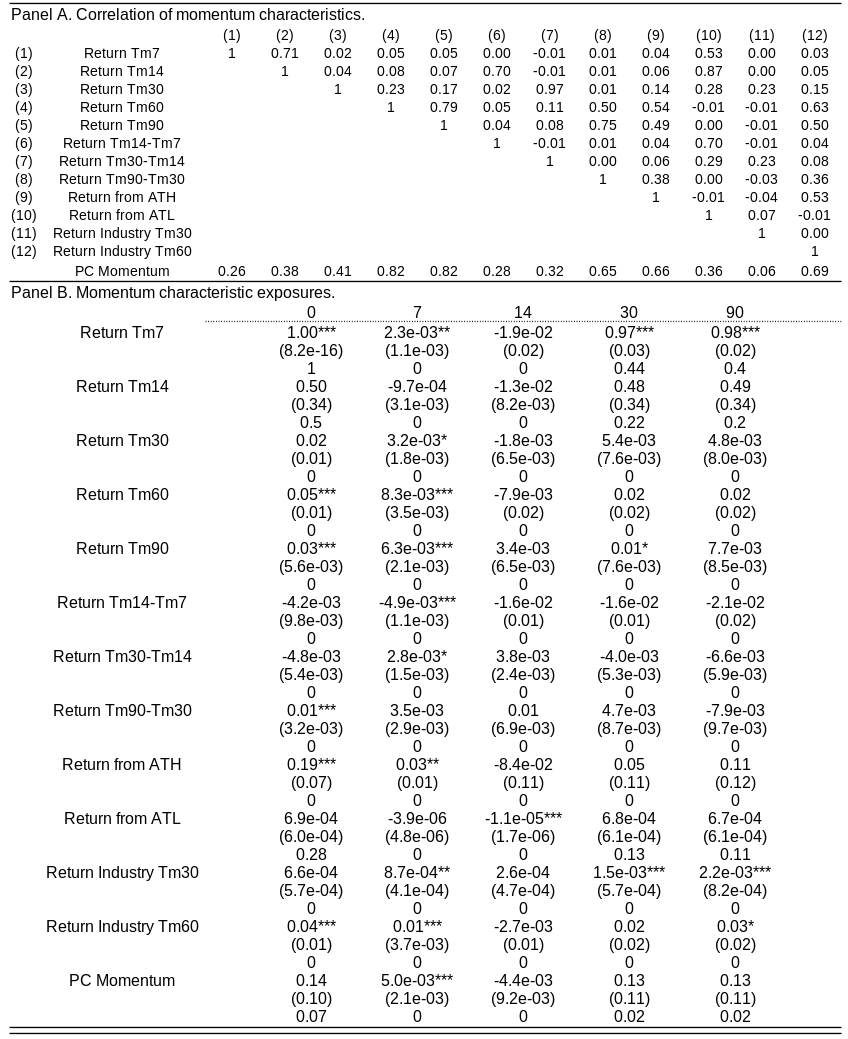}
\note{{\scriptsize This table reports the correlation matrix among Momentum Characteristics and the loadings on asset excess returns on each characteristic at various horizons. Panel A reports pairwise Pearson correlation coefficients among the characteristics and the first principal component from them. The characteristics are re-scaled to be mean zero and unit variance before PCA and studying these correlations. Panel B reports the coefficient (with 1, 2, and 3 stars for significant at the 10\%, 5\%, and 1\% levels, respectively), standard error, and $R^2$ for univariate panel regressions of asset excess weekly returns at 0, 7, 14, 30, and 90 days ahead on each of the characteristics and a constant. Standard errors are Newey-West adjusted using Bartlett's formula for the number of lags. There are 22,678 observations.}}
\label{f:momentum}\end{table}

\begin{table}[p]
\caption{Microstructure Characteristics: Correlations and Signal.}
\includegraphics[width=5.9in]{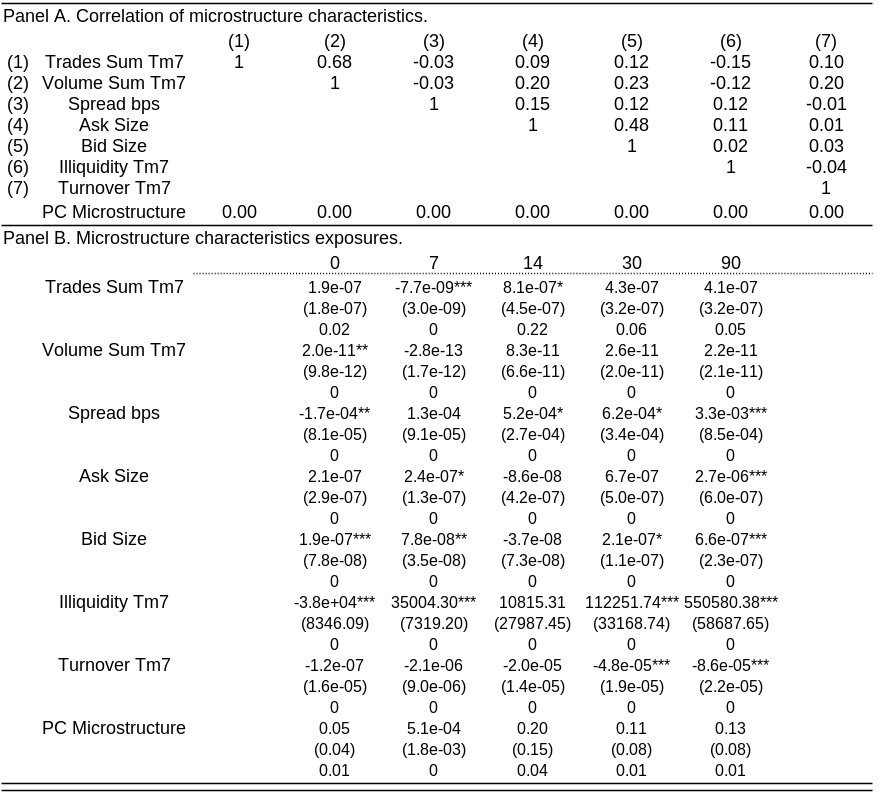}
\note{This table reports the correlation matrix among Microstructure Characteristics and the loadings on asset excess returns on each characteristic at various horizons. Panel A reports pairwise Pearson correlation coefficients among the characteristics and the first principal component from from them. The characteristics are re-scaled to be mean zero and unit variance before PCA and studying these correlations. Panel B reports the coefficient (with 1, 2, and 3 stars for significant at the 10\%, 5\%, and 1\% levels, respectively), standard error, and $R^2$ for univariate panel regressions of asset excess weekly returns at 0, 7, 14, 30, and 90 days ahead on each of the characteristics and a constant. Standard errors are Newey-West adjusted using Bartlett's formula for the number of lags. There are 22,678 observations.}
\label{f:microstructure}\end{table}

\begin{landscape}
\begin{table}[p]
\caption{Financial Characteristics: Correlations.}
\includegraphics[width=8.5in]{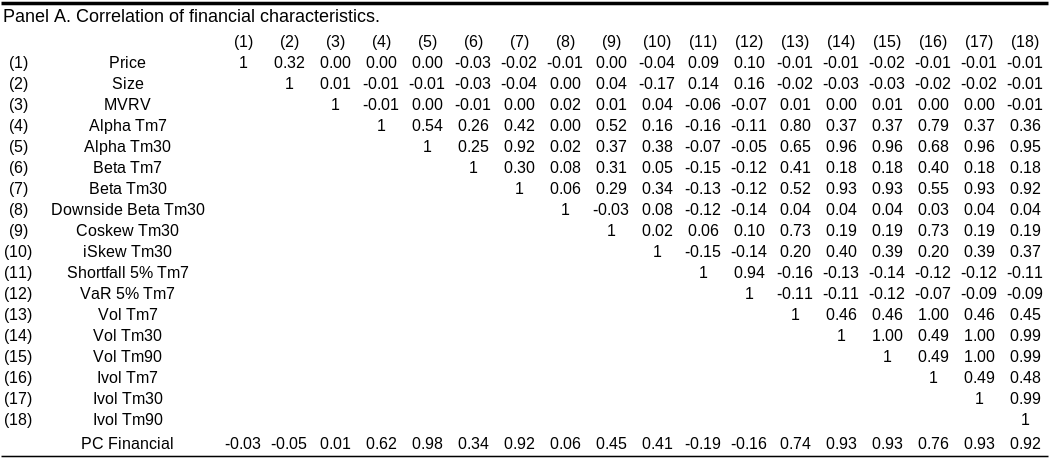}
\begin{minipage}{8.5in} 
This table reports the correlation matrix among Financial Characteristics. Panel A reports pairwise Pearson correlation coefficients among the characteristics and the first principal component from them. The characteristics are re-scaled to be mean zero and unit variance before PCA and studying these correlations.
\end{minipage}
\label{f:financial_panela}\end{table}
\end{landscape}

\begin{table}[p]
\caption{Financial Characteristics: Signal.}
\includegraphics[width=\textwidth]{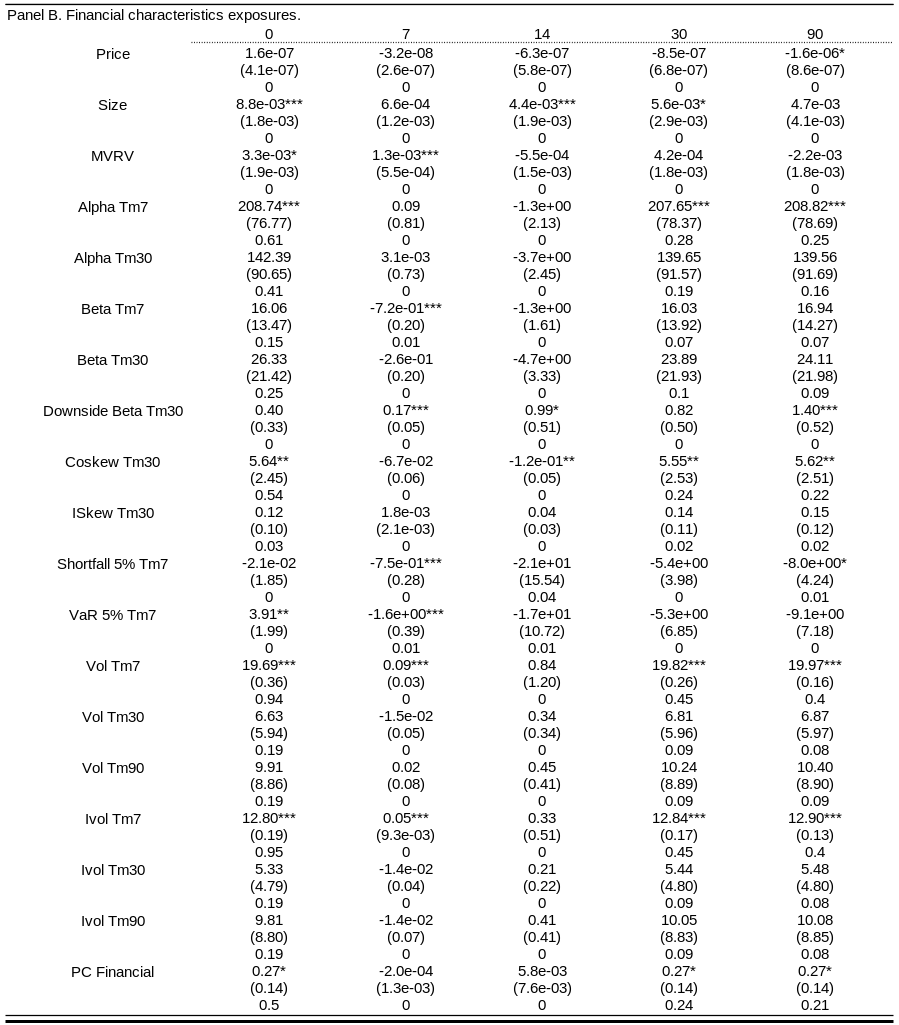}
\note{This table reports the loadings on asset excess returns on each characteristic at various horizons. Panel B reports the coefficient (with 1, 2, and 3 stars for significant at the 10\%, 5\%, and 1\% levels, respectively), standard error, and $R^2$ for univariate panel regressions of asset excess weekly returns at 0, 7, 14, 30, and 90 days ahead on each of the characteristics and a constant. Standard errors are Newey-West adjusted using Bartlett's formula for the number of lags. There are 22,678 observations.}
\label{f:financial_panelb}\end{table}

\begin{table}[p]
\caption{Principal Components of Characteristics: Correlations.}
\includegraphics[width=4in]{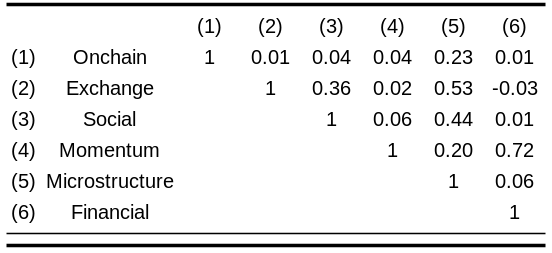}
\note{This table reports the correlation matrix among the first principal components of all groupings of asset characteristics, i.e. all pairwise Pearson correlation coefficients.}
\label{f:corr_pc}\end{table}

\begin{table}[p]
\caption{Characteristic Signal by Year.}
\includegraphics[width=\textwidth]{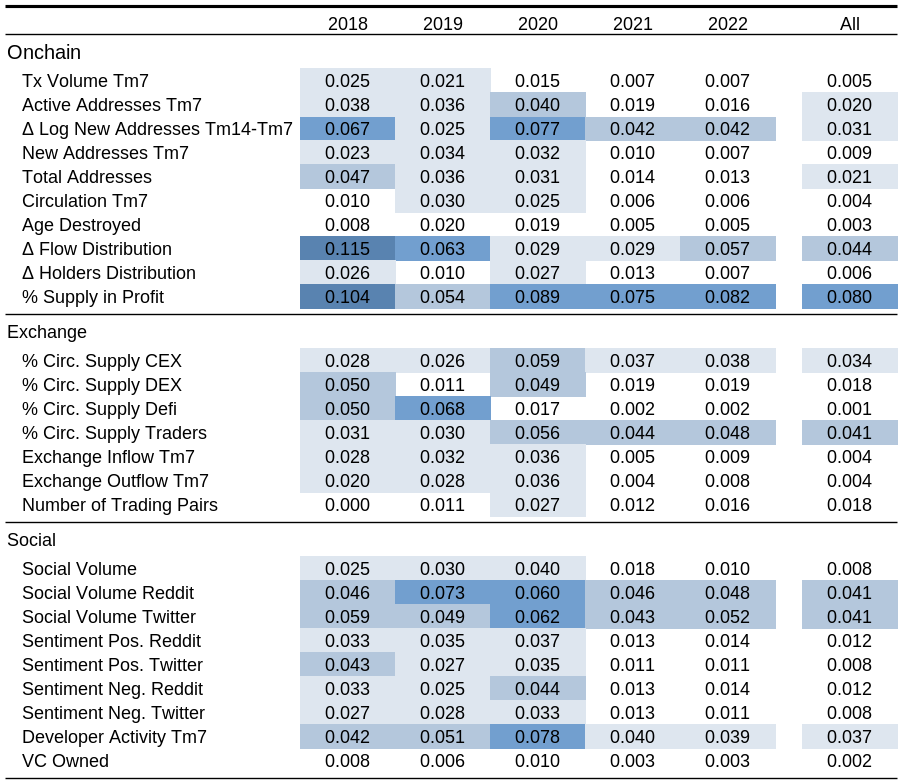}
\note{This table reports, by year and overall, the pairwise mutual information between all weekly panel characteristics and asset excess returns seven days ahead.}
\label{f:mi_panela}\end{table}

\begin{table}[p]
\caption{Characteristic Signal by Year (Continued).}
\includegraphics[width=\textwidth]{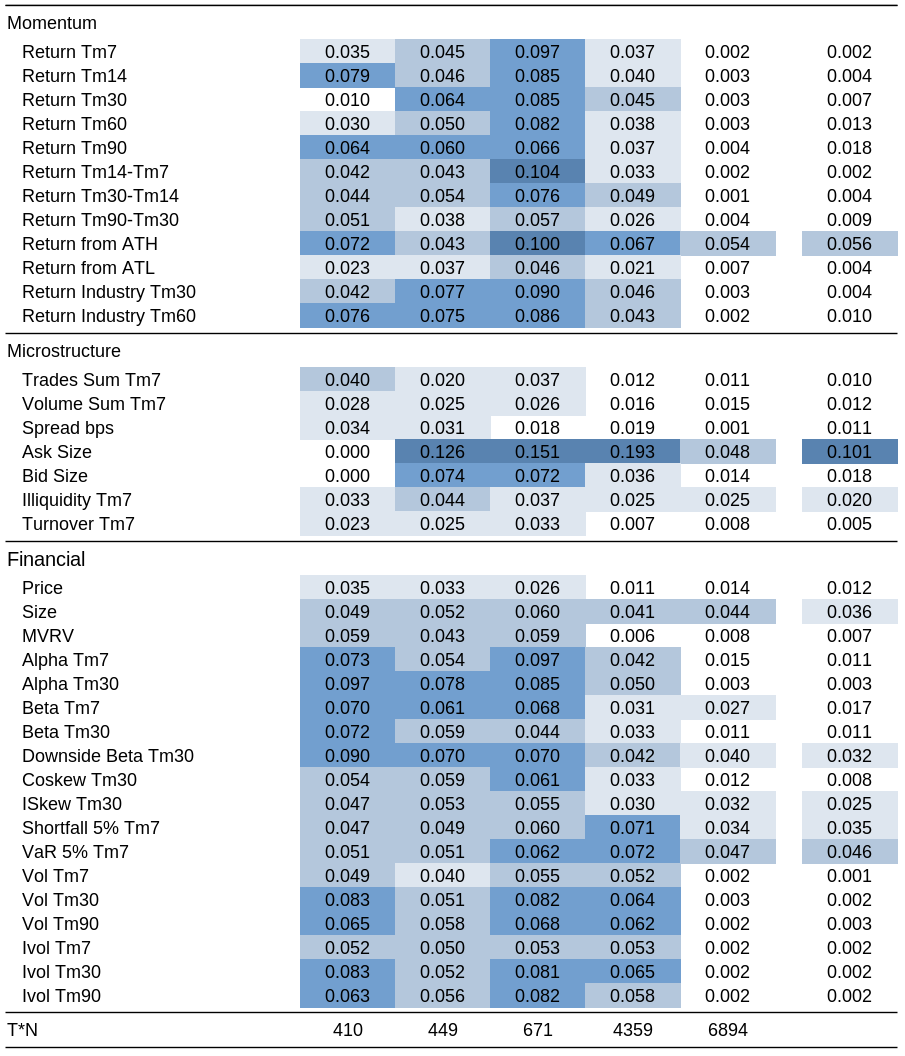}
\note{This table reports, by year and overall, the pairwise mutual information between all weekly panel characteristics and asset excess returns seven days ahead.}
\label{f:mi_panelb}\end{table}

\begin{table}[p]
\caption{Univariate Factor Returns: Statistically Significant Strategies.}
\includegraphics[width=\textwidth]{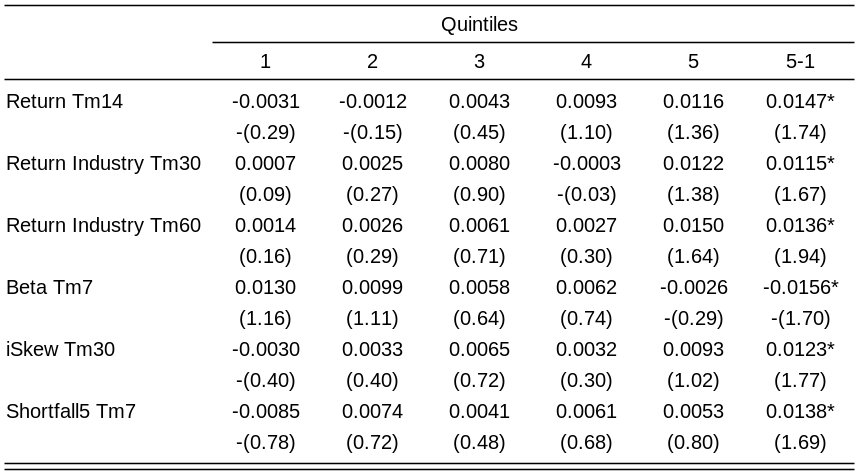}
\note{This table reports the mean quintile portfolio returns (and t-statistics) for characteristics with significant zero-investment strategies. The mean returns are the time-series averages of weekly value-weighted portfolio excess returns. 5-1 is the long-short top minus bottom quintile zero-investment portfolio. *, **, and *** denote significance at the 10\%, 5\%, and 1\% levels.}
\label{f:uni_factors_sig}\end{table}

\begin{table}[p]
\caption{Univariate Factor Returns: Onchain Strategies.}
\includegraphics[width=\textwidth]{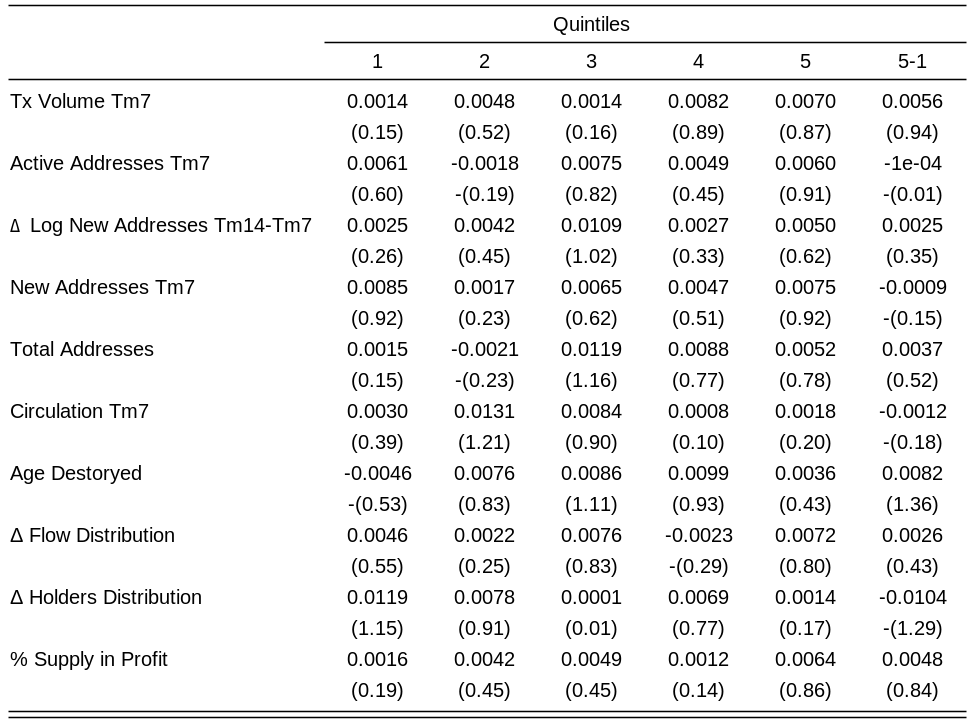}
\note{This table reports the mean quintile portfolio returns (and t-statistics) for onchain characteristics. The mean returns are the time-series averages of weekly value-weighted portfolio excess returns. 5-1 is the long-short top minus bottom quintile zero-investment portfolio. *, **, and *** denote significance at the 10\%, 5\%, and 1\% levels.}
\label{f:uni_factors_onchain}\end{table}

\begin{table}[p]
\caption{Univariate Factor Returns: Exchange Strategies.}
\includegraphics[width=\textwidth]{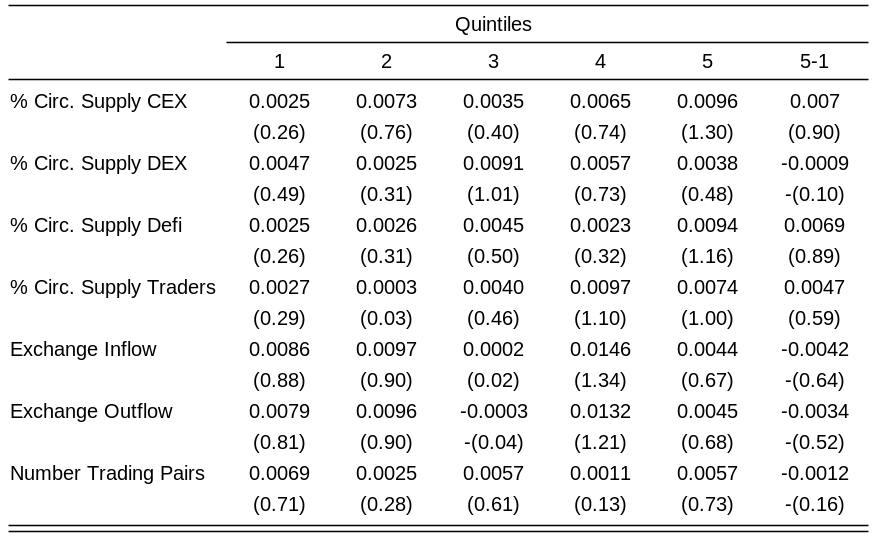}
\note{This table reports the mean quintile sorted portfolio returns (and t-statistics) for exchange characteristics. The mean returns are the time-series averages of weekly value-weighted portfolio excess returns. 5-1 is the long-short top minus bottom quintile zero-investment portfolio. *, **, and *** denote significance at the 10\%, 5\%, and 1\% levels.}
\label{f:uni_factors_exchange}\end{table}

\begin{table}[p]
\caption{Univariate Factor Returns: Social Strategies.}
\includegraphics[width=\textwidth]{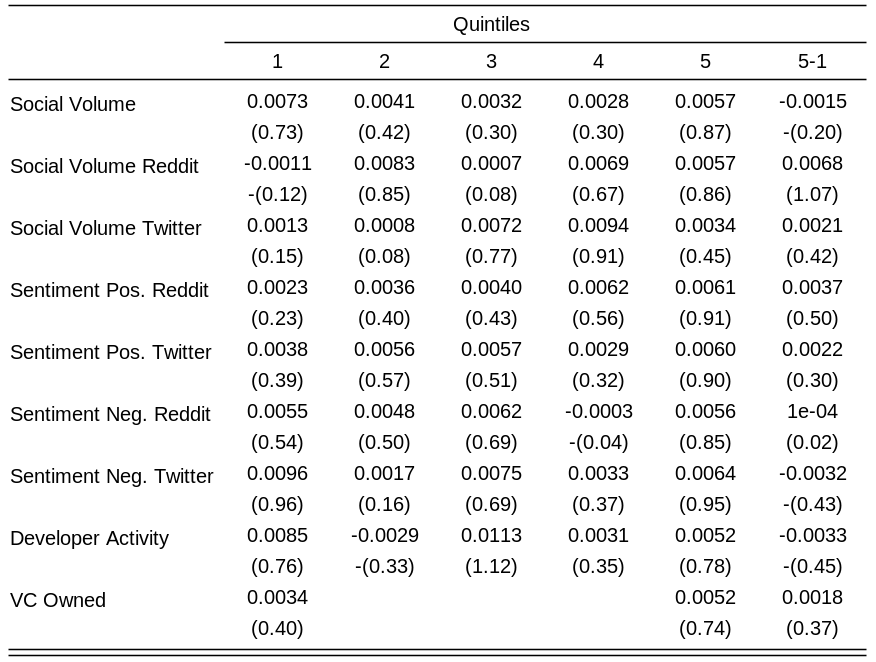}
\note{This table reports the mean quintile portfolio returns (and t-statistics) for social characteristics. The mean returns are the time-series averages of weekly value-weighted portfolio excess returns. 5-1 is the long-short top minus bottom quintile zero-investment portfolio. *, **, and *** denote significance at the 10\%, 5\%, and 1\% levels.}
\label{f:uni_factors_social}\end{table}

\begin{table}[p]
\caption{Univariate Factor Returns: Momentum Strategies.}
\includegraphics[width=\textwidth]{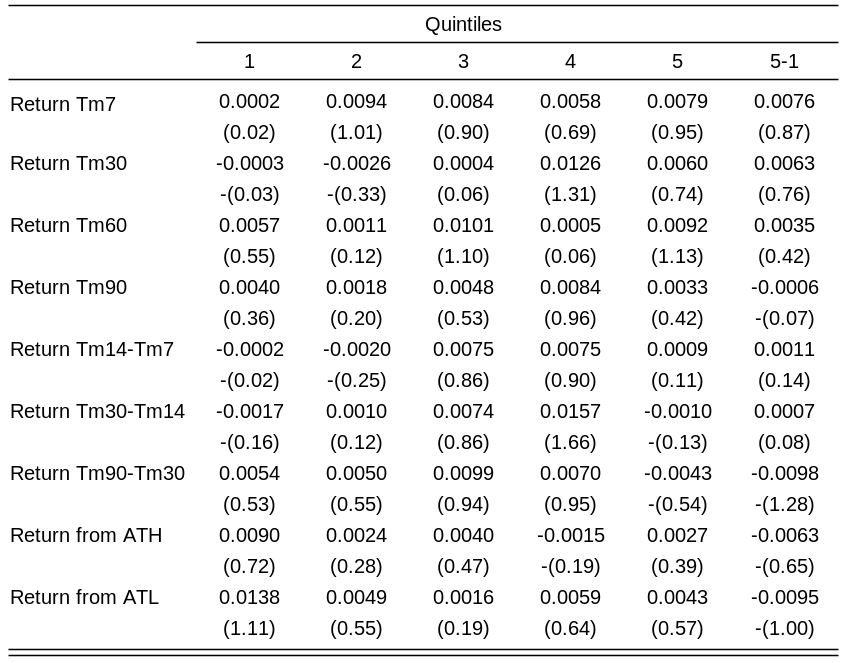}
\note{This table reports the mean quintile portfolio returns (and t-statistics) for momentum characteristics. The mean returns are the time-series averages of weekly value-weighted portfolio excess returns. 5-1 is the long-short top minus bottom quintile zero-investment portfolio. *, **, and *** denote significance at the 10\%, 5\%, and 1\% levels.}
\label{f:uni_factors_mom}\end{table}

\begin{table}[p]
\caption{Univariate Factor Returns: Microstructure Strategies.}
\includegraphics[width=\textwidth]{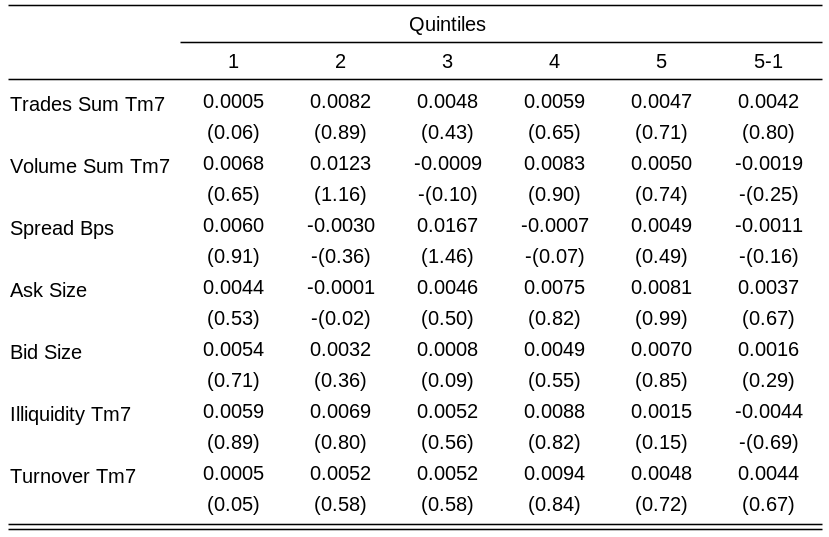}
\note{This table reports the mean quintile portfolio returns (and t-statistics) for microstructure characteristics. The mean returns are the time-series averages of weekly value-weighted portfolio excess returns. 5-1 is the long-short top minus bottom quintile zero-investment portfolio. *, **, and *** denote significance at the 10\%, 5\%, and 1\% levels.}
\label{f:uni_factors_micro}\end{table}

\begin{table}[p]
\caption{Univariate Factor Returns: Financial Strategies.}
\includegraphics[width=\textwidth]{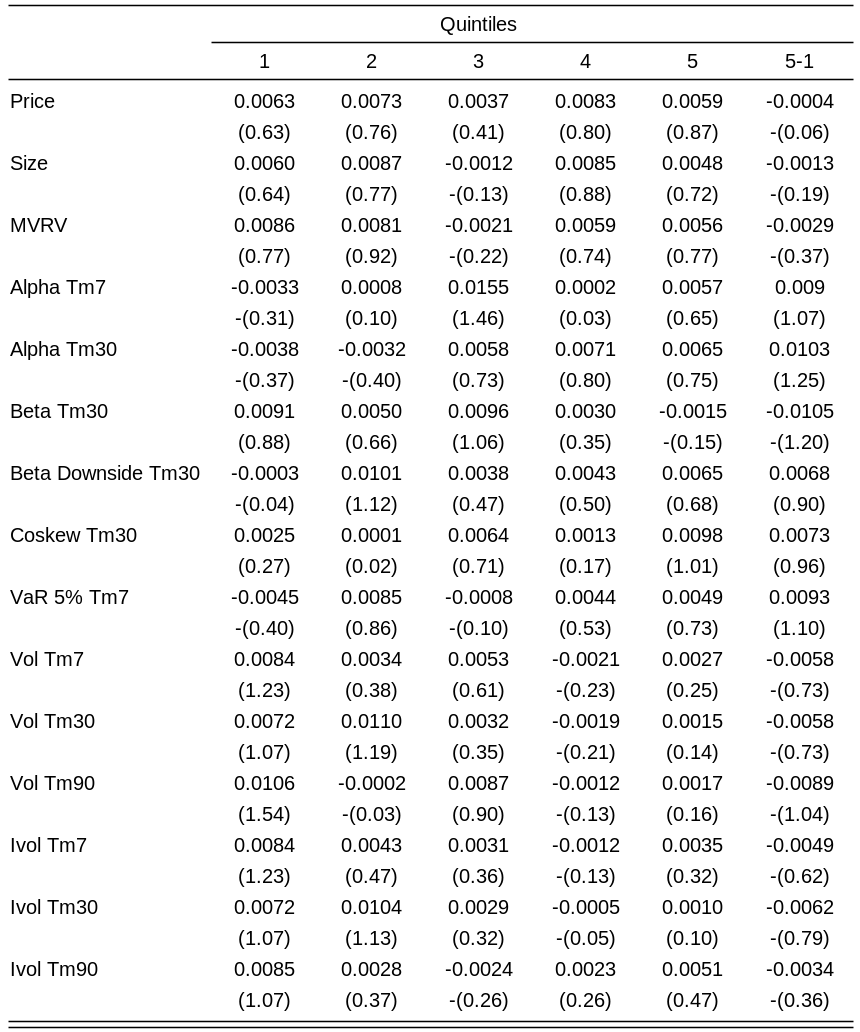}
\note{This table reports the mean quintile portfolio returns (and t-statistics) for financial characteristics. The mean returns are the time-series averages of weekly value-weighted portfolio excess returns. 5-1 is the long-short top minus bottom quintile zero-investment portfolio. *, **, and *** denote significance at the 10\%, 5\%, and 1\% levels.}
\label{f:uni_factors_fin}\end{table}

\end{document}